\pdfoutput=1

\documentclass[12pt,a4paper]{article}

\usepackage{ifthen} 
\newboolean{pdflatex}
\setboolean{pdflatex}{true} 

\newboolean{articletitles}
\setboolean{articletitles}{true} 

\newboolean{uprightparticles}
\setboolean{uprightparticles}{false} 

\newboolean{inbibliography}
\setboolean{inbibliography}{false} 

\usepackage{rotating}
\usepackage{multirow}
\usepackage{geometry}
\usepackage{afterpage} 

\usepackage{microtype}
\usepackage{lineno}  
\usepackage{xspace} 
\usepackage{caption} 

\usepackage{graphicx}  
\usepackage{color}
\usepackage{colortbl}
\graphicspath{{./figs/}} 

\usepackage{amsmath} 
\usepackage{amssymb}
\usepackage{amsfonts}
\usepackage{upgreek} 

\newcommand*\patchAmsMathEnvironmentForLineno[1]{%
\expandafter\let\csname old#1\expandafter\endcsname\csname #1\endcsname
\expandafter\let\csname oldend#1\expandafter\endcsname\csname
end#1\endcsname
 \renewenvironment{#1}%
   {\linenomath\csname old#1\endcsname}%
   {\csname oldend#1\endcsname\endlinenomath}%
}
\newcommand*\patchBothAmsMathEnvironmentsForLineno[1]{%
  \patchAmsMathEnvironmentForLineno{#1}%
  \patchAmsMathEnvironmentForLineno{#1*}%
}
\AtBeginDocument{%
\patchBothAmsMathEnvironmentsForLineno{equation}%
\patchBothAmsMathEnvironmentsForLineno{align}%
\patchBothAmsMathEnvironmentsForLineno{flalign}%
\patchBothAmsMathEnvironmentsForLineno{alignat}%
\patchBothAmsMathEnvironmentsForLineno{gather}%
\patchBothAmsMathEnvironmentsForLineno{multline}%
}

\usepackage{hyperref}    
\usepackage[all]{hypcap} 

\def\lhcb {\mbox{LHCb}\xspace}

\ifthenelse{\boolean{uprightparticles}}%
{

 \def\Pmu         {\ensuremath{\upmu}\xspace}

 \def\Ppi         {\ensuremath{\uppi}\xspace}

 \def\Ppsi        {\ensuremath{\uppsi}\xspace}

 \def\PDelta      {\ensuremath{\Delta}\xspace}                 
 \def\PXi      {\ensuremath{\Xi}\xspace}                 
 \def\PLambda      {\ensuremath{\Lambda}\xspace}                 
 \def\PSigma      {\ensuremath{\Sigma}\xspace}                 
 \def\POmega      {\ensuremath{\Omega}\xspace}                 
 \def\PUpsilon      {\ensuremath{\Upsilon}\xspace}                 
 
 \def\PB      {\ensuremath{\mathrm{B}}\xspace}                 
                  
 \def\PD      {\ensuremath{\mathrm{D}}\xspace}

 \def\PJ      {\ensuremath{\mathrm{J}}\xspace}                 
 \def\PK      {\ensuremath{\mathrm{K}}\xspace}

 \def\Pb      {\ensuremath{\mathrm{b}}\xspace}                 
 \def\Pc      {\ensuremath{\mathrm{c}}\xspace}

 \def\Pi      {\ensuremath{\mathrm{i}}\xspace}

}
{

 \def\Pmu         {\ensuremath{\mu}\xspace}

 \def\Ppi         {\ensuremath{\pi}\xspace}

 \def\Ppsi        {\ensuremath{\psi}\xspace}                 
                  
 \mathchardef\PDelta="7101
 \mathchardef\PXi="7104
 \mathchardef\PLambda="7103
 \mathchardef\PSigma="7106
 \mathchardef\POmega="710A
 \mathchardef\PUpsilon="7107
                  
 \def\PB      {\ensuremath{B}\xspace}                 
                  
 \def\PD      {\ensuremath{D}\xspace}

 \def\PJ      {\ensuremath{J}\xspace}                 
 \def\PK      {\ensuremath{K}\xspace}

 \def\Pb      {\ensuremath{b}\xspace}                 
 \def\Pc      {\ensuremath{c}\xspace}

 \def\Pi      {\ensuremath{i}\xspace}

}

\def\mumu       {\ensuremath{\Pmu^+\Pmu^-}\xspace}

\def\cquark    {\ensuremath{\Pc}\xspace}

\def\bquark    {\ensuremath{\Pb}\xspace}

\def\pion  {\ensuremath{\Ppi}\xspace}

\def\pip   {\ensuremath{\pion^+}\xspace}
\def\pim   {\ensuremath{\pion^-}\xspace}

\def\kaon  {\ensuremath{\PK}\xspace}
  \def\Kbar  {\kern 0.2em\overline{\kern -0.2em \PK}{}\xspace}

\def\Kp    {\ensuremath{\kaon^+}\xspace}
\def\Km    {\ensuremath{\kaon^-}\xspace}

\def\Kstarz  {\ensuremath{\kaon^{*0}}\xspace}

  \def\Dbar    {\kern 0.2em\overline{\kern -0.2em \PD}{}\xspace}
\def\D       {\ensuremath{\PD}\xspace}

\def\Dzb     {\ensuremath{\Dbar^0}\xspace}

\def\Dm      {\ensuremath{\D^-}\xspace}

\def\B       {\ensuremath{\PB}\xspace}
\def\Bbar    {\ensuremath{\kern 0.18em\overline{\kern -0.18em \PB}{}}\xspace}

\def\Bz      {\ensuremath{\B^0}\xspace}

\def\Bu      {\ensuremath{\B^+}\xspace}

\def\Bp      {\ensuremath{\Bu}\xspace}

\def\jpsi     {\ensuremath{{\PJ\mskip -3mu/\mskip -2mu\Ppsi\mskip 2mu}}\xspace}

  \def\Y#1S{\ensuremath{\PUpsilon{(#1S)}}\xspace}

\def\Lbar {\ensuremath{\kern 0.1em\overline{\kern -0.1em\PLambda}}\xspace}

\def\to                 {\ensuremath{\rightarrow}\xspace}

\def\AT#1     {\ensuremath{A_{\mathrm{T}}^{#1}}\xspace}           

\def\C#1      {\ensuremath{\mathcal{C}_{#1}}\xspace}                       
\def\Cp#1     {\ensuremath{\mathcal{C}_{#1}^{'}}\xspace}                    
\def\Ceff#1   {\ensuremath{\mathcal{C}_{#1}^{\mathrm{(eff)}}}\xspace}        
\def\Cpeff#1  {\ensuremath{\mathcal{C}_{#1}^{'\mathrm{(eff)}}}\xspace}       
\def\Ope#1    {\ensuremath{\mathcal{O}_{#1}}\xspace}                       
\def\Opep#1   {\ensuremath{\mathcal{O}_{#1}^{'}}\xspace}                    

\newcommand{\tev}{\ifthenelse{\boolean{inbibliography}}{\ensuremath{~T\kern -0.05em eV}\xspace}{\ensuremath{\mathrm{\,Te\kern -0.1em V}}\xspace}}
\newcommand{\gev}{\ensuremath{\mathrm{\,Ge\kern -0.1em V}}\xspace}
\newcommand{\mev}{\ensuremath{\mathrm{\,Me\kern -0.1em V}}\xspace}
\newcommand{\kev}{\ensuremath{\mathrm{\,ke\kern -0.1em V}}\xspace}
\newcommand{\ev}{\ensuremath{\mathrm{\,e\kern -0.1em V}}\xspace}
\newcommand{\gevc}{\ensuremath{{\mathrm{\,Ge\kern -0.1em V\!/}c}}\xspace}
\newcommand{\mevc}{\ensuremath{{\mathrm{\,Me\kern -0.1em V\!/}c}}\xspace}
\newcommand{\gevcc}{\ensuremath{{\mathrm{\,Ge\kern -0.1em V\!/}c^2}}\xspace}
\newcommand{\gevgevcccc}{\ensuremath{{\mathrm{\,Ge\kern -0.1em V^2\!/}c^4}}\xspace}
\newcommand{\mevcc}{\ensuremath{{\mathrm{\,Me\kern -0.1em V\!/}c^2}}\xspace}

\def\mum  {\ensuremath{{\,\upmu\rm m}}\xspace}

\def\invfb   {\ensuremath{\mbox{\,fb}^{-1}}\xspace}

\def\ps   {\ensuremath{{\rm \,ps}}\xspace}

\newcommand{\chisq}{\ensuremath{\chi^2}\xspace}
\newcommand{\chisqndf}{\ensuremath{\chi^2/\mathrm{ndf}}\xspace}

\def\gsim{{~\raise.15em\hbox{$>$}\kern-.85em
          \lower.35em\hbox{$\sim$}~}\xspace}
\def\lsim{{~\raise.15em\hbox{$<$}\kern-.85em
          \lower.35em\hbox{$\sim$}~}\xspace}

\def\pt         {\mbox{$p_{\rm T}$}\xspace}

\def\evtgen     {\mbox{\textsc{EvtGen}}\xspace}

\def\geant      {\mbox{\textsc{Geant4}}\xspace}

\def\photos     {\mbox{\textsc{Photos}}\xspace}

\def\pythia     {\mbox{\textsc{Pythia}}\xspace}

\def\tell1  {TELL1\xspace}
\def\ukl1   {UKL1\xspace}

\newcommand{\eg}{\mbox{\itshape e.g.}\xspace}

\def\Bstar   {\ensuremath{\B^{*}}\xspace}
\def\Bus     {\ensuremath{\B^{*+}}\xspace}
\def\Bds     {\ensuremath{\B^{*0}}\xspace}
\def\Bss     {\ensuremath{\B^{*0}_s}\xspace}
\def\Bss     {\ensuremath{\B^{**}}\xspace}

\def\Bsss    {\ensuremath{\B^{**0}_s}\xspace}

\usepackage{cite} 
\usepackage{mciteplus}

\def\phani{\phantom{1}}
\def\phanii{\phantom{11}}
\def\phanm{\phantom{-}}
\def\phand{\phantom{.}}
\def\phanid{\phantom{1.}}

\begin{document}

\renewcommand{\thefootnote}{\fnsymbol{footnote}}
\setcounter{footnote}{1}


\begin{titlepage}
\pagenumbering{roman}

\vspace*{-1.5cm}
\centerline{\large EUROPEAN ORGANIZATION FOR NUCLEAR RESEARCH (CERN)}
\vspace*{1.5cm}
\hspace*{-0.5cm}
\begin{tabular*}{\linewidth}{lc@{\extracolsep{\fill}}r}
\ifthenelse{\boolean{pdflatex}}
{\vspace*{-2.7cm}\mbox{\!\!\!\includegraphics[width=.14\textwidth]{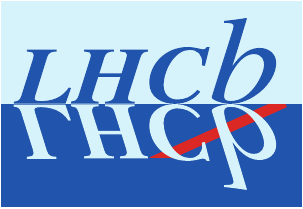}} & &}%
{\vspace*{-1.2cm}\mbox{\!\!\!\includegraphics[width=.12\textwidth]{lhcb-logo.eps}} & &}%
\\
 & & CERN-PH-EP-2015-021 \\  
 & & LHCb-PAPER-2014-067 \\  
 & & \today \\ 
 & & \\
\end{tabular*}

\vspace*{2.0cm}

{\bf\boldmath\huge
\begin{center}
  Precise measurements of the properties of the $B_1(5721)^{0,+}$ and $B_2^*(5747)^{0,+}$ states and 
  observation of $B^{+,0} \pi^{-,+}$ mass structures
\end{center}
}

\vspace*{1.0cm}

\begin{center}
The LHCb collaboration\footnote{Authors are listed at the end of this paper.}
\end{center}

\vspace{\fill}

\begin{abstract}
  \noindent
  Invariant mass distributions of $\Bp\pim$ and $\Bz\pip$ combinations are investigated in order to study excited \B mesons. 
  The analysis is based on a data sample corresponding to $3.0 \invfb$ of $pp$ collision data, recorded by the \lhcb detector at centre-of-mass energies of 7 and $8 \tev$. 
  Precise measurements of the masses and widths of the $B_1(5721)^{0,+}$ and $B_2^*(5747)^{0,+}$ states are reported.
  Clear enhancements, particularly prominent at high pion transverse momentum, are seen over background in the mass range $5850$--$6000 \mev$ in both $\Bp\pim$ and $\Bz\pip$ combinations.
The structures are consistent with the presence of four excited \B mesons, labelled $B_J(5840)^{0,+}$ and $B_J(5960)^{0,+}$, whose masses and widths are obtained under different hypotheses for their quantum numbers.
\end{abstract}

\vspace*{1.0cm}

\begin{center}
  Published in JHEP
\end{center}

\vspace{\fill}

{\footnotesize 
\centerline{\copyright~CERN on behalf of the \lhcb collaboration, licence \href{http://creativecommons.org/licenses/by/4.0/}{CC-BY-4.0}.}}
\vspace*{2mm}

\end{titlepage}


\newpage
\setcounter{page}{2}
\mbox{~}

\cleardoublepage


\renewcommand{\thefootnote}{\arabic{footnote}}
\setcounter{footnote}{0}



\pagestyle{plain} 
\setcounter{page}{1}
\pagenumbering{arabic}


%

\section{Introduction}
\label{sec:Introduction}

The properties of excited $B$ mesons containing a light quark can be described in the context of heavy quark effective theory (HQET)~\cite{Mannel:1996cn}. 
Since the mass of the $b$ quark is much larger than the QCD scale, the Lagrangian can be expanded in powers of $1/m_{b}$, where the leading term defines the static limit ($m_b  \to \infty $).
In the heavy quark approximation, the \B mesons 
are characterised by three quantum numbers: 
the orbital angular momentum $L$ (S, P, D for $L = 0, 1, 2$ respectively) of the light quark,
its total angular
momentum  $j_q = |L \pm  \frac{1}{2}|$, and the total angular momentum $J = |j_q \pm \frac{1}{2}|$ of the \B meson. 
 The spectroscopic notation has the form $n^{2S+1}L_J$, where $S=0$ or $1$ is the sum of the
 quark spins and where the quantum number $n$  describes the radial excitations of the state. 
The PDG notation~\cite{PDG2014} (which is used in this paper) has the form $B_{J}^{(*)}(m)$ or $B_{J}^{(*)}(nL)$, where $m$ is the mass in units of \mev,\footnote{Natural units where {\it $\hbar$} = {\it c} = 1 are used.} the $*$ superscript is given to 
those states with natural spin-parity $P = (-1)^J$
($J^P = 0^+, 1^-, 2^+, ...$),
and the subscript $J$ is omitted for pseudoscalar and vector states. A prime may be used to distinguish two states with the same quantum numbers.

\begin{figure}[!tb]
  \centering
  \includegraphics[width=0.7\textwidth]{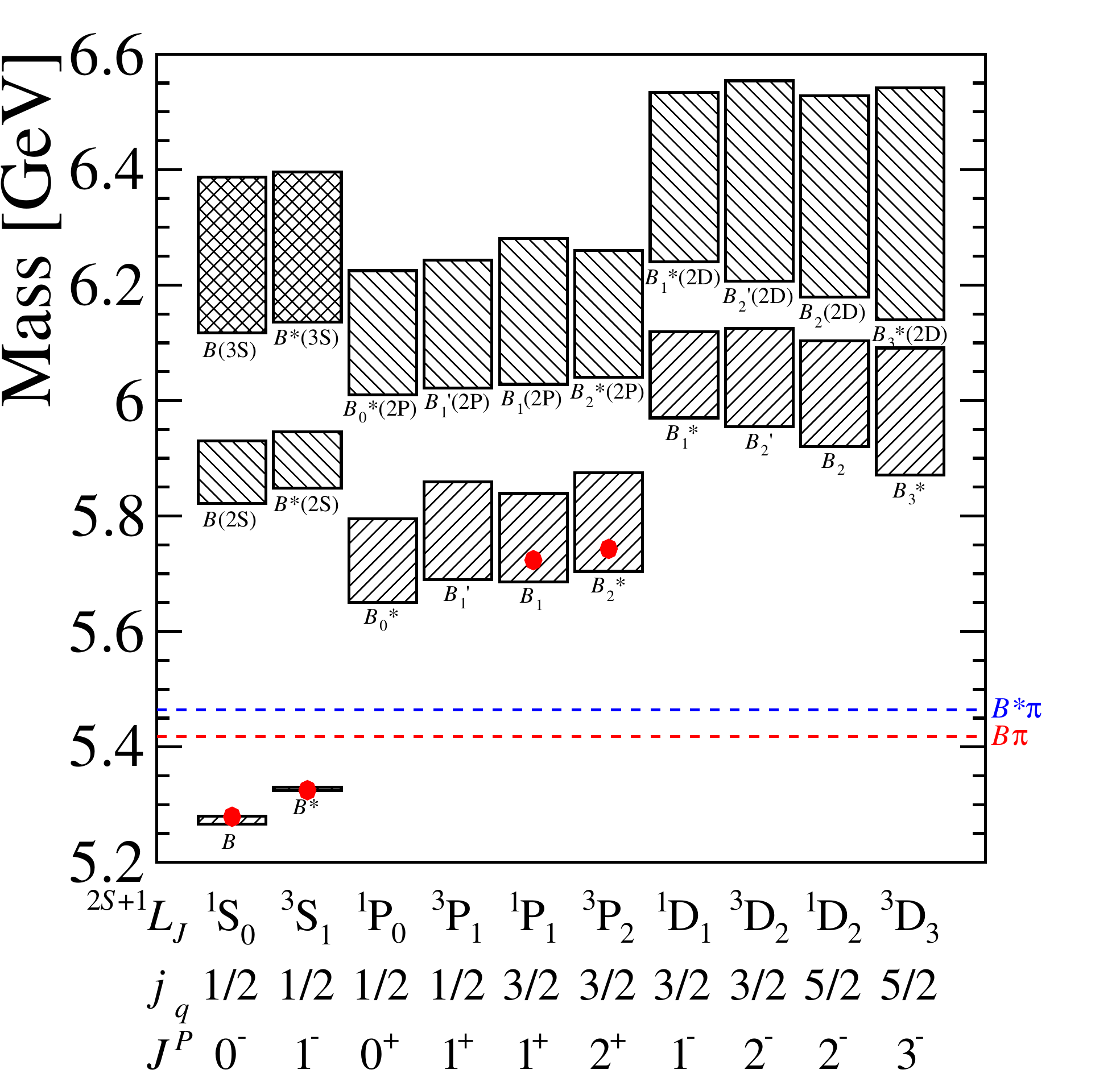}
  \caption{
    Mass predictions of the excited \B states~\cite{DiPierro:2001uu, Ebert:2009ua, Zeng:1994vj, Gupta:1994mw, Lahde:1999ih, Dai:1993np, Devlani:2012zz, Colangelo:2012xi}.
    The boxes cover the range of predictions for the masses of each state, and the red dots indicate the measured values. 
    The horizontal lines correspond to the $B\pi$ (red) and $B^*\pi$ (blue) thresholds. 
    \label{fig:B_spectrum} 
  }
\end{figure}

For $L = 0$, there are two possible ($J$; $j_q$) combinations, both
parity-odd, corresponding to the \B meson ground state with $J^P = 0^-$ and
to the excited $B^*$ state with $J^P = 1^-$. 
Higher excitations are collectively referred to as \Bss states and decay
strongly to lighter \B mesons and pions.
For $L = 1$ there are four different possible ($J$; $j_q$) combinations, all parity-even.
Predictions for the masses of such states and higher excitations spread over a wide range
of values, as shown in Fig.~\ref{fig:B_spectrum}~\cite{DiPierro:2001uu, Ebert:2009ua, Zeng:1994vj, Gupta:1994mw, Lahde:1999ih, Dai:1993np, Devlani:2012zz, Colangelo:2012xi}.
As can be seen in Fig.~\ref{fig:B_spectrum}, the states come in doublets (two values of $J$ for each $j_q$), and within each doublet, one has natural 
and one unnatural spin-parity quantum numbers.
States with natural spin-parity (except for $0^+$) can decay to both $B\pi$ and $B^*\pi$ final states.  
States with unnatural spin-parity cannot decay to the pseudoscalar-pseudoscalar $B\pi$ final state due to parity conservation, but may decay to $B^*\pi$ (Table~\ref{tab:bstst}).
Since the $B^*$ meson decays to $B\gamma$, the signature from a doublet of \Bss states is given by three peaks in the $B\pi$ mass spectrum (unless the doublet includes a $0^+$ state): one from the natural spin-parity state decay to $B\pi$, and two from both states decaying to $B^*\pi$ with a missing photon.
Due to the missing photon, the peaks from $B^*\pi$ decays are shifted down from the true \Bss mass by the difference between the $B^*$ and $B$ masses (this feature recently allowed a precise determination of the $B^*-B$ mass difference from the $\Bp\Km$ spectrum~\cite{LHCb-PAPER-2012-030}).
Depending on the widths of the states and the mass resolution, two or all three of these peaks may overlap and be hard to distinguish experimentally.
The $B_0^*$ and $B_1^\prime$ states are predicted to be very broad~\cite{DiPierro:2001uu, Colangelo:2012xi}
 since they decay via S-wave (the comparable states in the charm sector have widths of around $300 \mev$~\cite{PDG2014}).
However, the $B_1$ and $B_2^*$ states decay only via D-wave and are predicted~\cite{DiPierro:2001uu, Colangelo:2012xi} and observed~\cite{PDG2014} to be much narrower.  
Higher states such as the $B(2{\rm S})$, $B^*(2{\rm S})$, $B_2(1{\rm D})$ and $B_3^*(1{\rm D})$ are predicted to have widths in the $100$--$200 \mev$ range~\cite{Colangelo:2012xi}, consistent with the recent measurement of the properties of the $D_{s3}^*(1{\rm D})$ state~\cite{LHCb-PAPER-2014-035,LHCb-PAPER-2014-036}.

\begin{table}[hbt]
\begin{center}
\caption{
  Allowed decay modes for the excited \B states.}
\begin{tabular}{ccccc}
\hline 
$J^P$ & \multicolumn{2}{c}{Allowed decay mode}\\ 
& \phantom{kk} $\B\pi$ \phantom{kk} & \phantom{kk} $B^* \pi$ \phantom{kk} \\
\hline
$0^+$ & yes & no \\
$0^-, 1^+, 2^-, ...$ & no & yes \\
$1^-, 2^+, 3^-, ...$ & yes & yes \\
\hline 
\end{tabular}
\label{tab:bstst}
\end{center}
\end{table}

In contrast to the situation in the charm sector,
there is relatively little experimental information concerning $B$ meson spectroscopy.
The $B_1(5721)^{0}$ and $B_2^*(5747)^{0}$ states have been observed by the CDF~\cite{:2008jn} and D0~\cite{Abazov:2007vq} experiments, 
and recently the CDF collaboration has presented results on the charged isospin partners, together with evidence for a higher mass resonance~\cite{Aaltonen:2013yya}.
This result has prompted theoretical speculation about the origin of the new state~\cite{Sun:2014wea,Wang:2014cta,Xu:2014mka,Gelhausen:2014jea,Xiao:2014ura}.
While in the \D meson system amplitude analyses of excited states produced in
\B decays can be used to determine their spin and parity (see, for example, Refs.~\cite{Abe:2003zm,LHCb-PAPER-2014-035,LHCb-PAPER-2014-036}), 
in the \B meson system it is very difficult to assign with certainty quantum numbers to observed states.
The labelling of the states follows the quark-model expectations for the
quantum numbers, which have not been experimentally verified.

In this paper, the results of a study of $\Bp\pim$ and $\Bz\pip$ combinations are presented. 
The inclusion of charge-conjugate processes is implied throughout.
The analysis is based on a data sample corresponding to $3.0 \invfb$ of LHC $pp$ collision data recorded with the LHCb detector at centre-of-mass energies of 7 and $8 \tev$. 

The $B$ mesons are reconstructed in the $\jpsi \Kp$, $\Dzb\pip$, $\Dzb\pip\pip\pim$, $\jpsi \Kstarz$, $\Dm\pip$ and $\Dm\pip\pip\pim$ channels, with subsequent $\jpsi \to \mumu$, $\Dzb \to \Kp\pim \ {\rm and} \ \Kp\pim\pip\pim$, $\Dm\to\Kp\pim\pim$ and $\Kstarz\to\Kp\pim$ decays.
The \B meson candidates are required to originate from a primary $pp$ collision vertex (PV), and are combined with pions originating from the same PV (referred to as ``companion pions'').
Both ``right-sign'' (RS) and ``wrong-sign'' (WS) combinations are considered, where the latter are those with quark-content that precludes that the pair originates from the strong decay of an excited $B$ meson (\eg $\Bp\pip$) and 
are used to model the combinatorial background.
Excited $B$ mesons are seen as peaks in the RS invariant mass distributions, and are fitted with relativistic Breit-Wigner (RBW) functions.
An additional very broad component, observed in the RS and not in the WS combinations, is referred to as ``associated production'' (AP) in this paper. 
The AP contribution may originate from very broad resonances or from correlated nonresonant production of $B$ mesons and companion pions in the fragmentation chain.

The remainder of the paper is organised as follows.
A brief description of the LHCb detector is given in Sec.~\ref{sec:Detector}.
The selection requirements are described in Sec.~\ref{sec:Selection},
the fit model is discussed in Sec.~\ref{sec:Fitmodel}, and the nominal fit results are given in Sec.~\ref{sec:nominalfit},
with the evaluation of the systematic uncertainties in Sec.~\ref{sec:systematics}. Interpretation of the results and a summary are given in Sec.~\ref{sec:interpretation}.

\section{Detector and dataset}
\label{sec:Detector}

The \lhcb detector~\cite{Alves:2008zz,LHCb-DP-2014-002} is a single-arm forward
spectrometer covering the \mbox{pseudorapidity} range $2<\eta <5$,
designed for the study of particles containing \bquark or \cquark
quarks. The detector includes a high-precision tracking system
consisting of a silicon-strip vertex detector~\cite{LHCb-DP-2014-001}
surrounding the $pp$ interaction region, a large-area silicon-strip detector
located upstream of a dipole magnet with a bending power of about
$4{\rm\,Tm}$, and three stations of silicon-strip detectors and straw
drift tubes~\cite{LHCb-DP-2013-003} placed downstream of the magnet.
The tracking system provides a momentum measurement with
relative uncertainty that varies from 0.5\% at low momentum to 1.0\% at
200\gev, and an impact parameter measurement with resolution of 20\mum for tracks with large momentum transverse to the beamline (\pt).
Different types of charged hadrons are distinguished using information
from two ring-imaging Cherenkov detectors~\cite{LHCb-DP-2012-003}. 
Photon, electron and hadron candidates are identified by a calorimeter system consisting of scintillating-pad and preshower detectors,
an electromagnetic calorimeter and a hadronic calorimeter. 
Muons are identified by a system composed of alternating layers of iron and multiwire proportional chambers~\cite{LHCb-DP-2012-002}.
The trigger~\cite{LHCb-DP-2012-004} consists of a hardware stage, based on information from the calorimeter and muon systems,
followed by a software stage, which uses information from the vertex detector
and tracking system.

In the simulation, $pp$ collisions are generated using
\pythia~\cite{Sjostrand:2006za} with a specific \lhcb
configuration~\cite{LHCb-PROC-2010-056}.  Decays of hadronic particles
are described by \evtgen~\cite{Lange:2001uf}, in which final-state
radiation is generated using \photos~\cite{Golonka:2005pn}. The
interaction of the generated particles with the detector, and its
response, are implemented using the \geant
toolkit~\cite{Allison:2006ve, *Agostinelli:2002hh} as described in
Ref.~\cite{LHCb-PROC-2011-006}.

\section{Event selection}
\label{sec:Selection}
The \B meson candidates in each decay mode are reconstructed using a set of loose selection requirements to suppress the majority of the combinatorial backgrounds. 
The selection criteria are similar to those used in previous analyses of the same channels~\cite{LHCb-PAPER-2013-065,LHCb-PAPER-2012-001,LHCb-PAPER-2011-016,LHCb-PAPER-2011-040}.
The $\Bu\to J/\psi K^+$ and  $\Bz\to J/\psi \Kstarz$  selections require a \B candidate with $\pt > 3\gev$ and a decay time of at least $0.3\ps$.
For the other decay modes, the selection explicitly requires that the software trigger decision 
 is based only on tracks from which the \B meson candidate is formed.
No requirement is imposed on how the event was selected at the hardware trigger stage.
Additional loose selection requirements are placed on variables related to the \B meson production and decay, such as transverse momentum and quality of the track fits for the decay products, detachment of the \B candidate from the PV, and whether the momentum of the \B candidate points back to the PV. 
Because \Bz mesons oscillate, the distinction between RS
and WS combinations is clearest at short \Bz decay times, and hence only \Bz candidates with decay time below $2\ps$ are used in the analysis.

\begin{figure*}[!tb]
 \centering
\includegraphics[width=.45\textwidth]{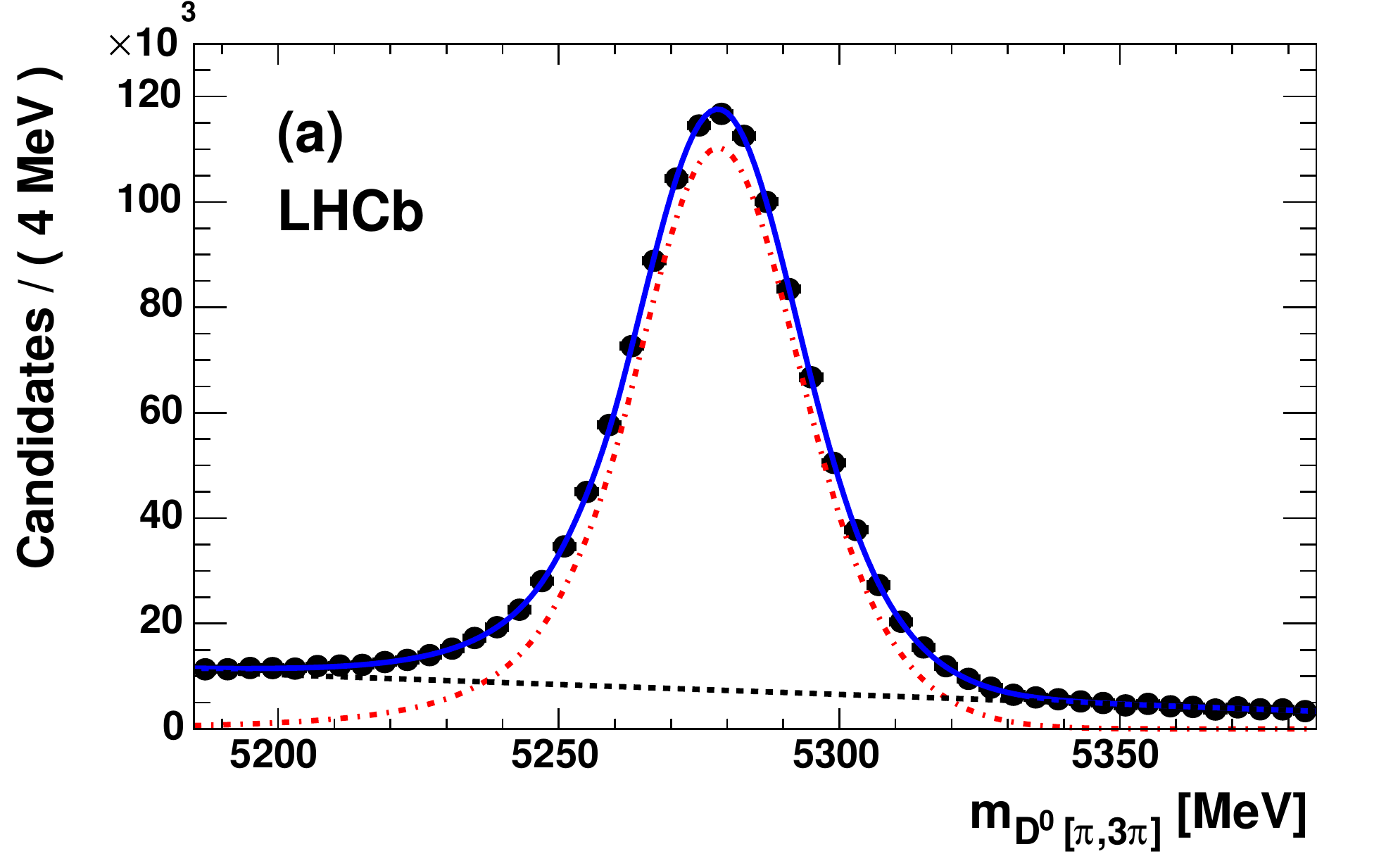}
\includegraphics[width=.45\textwidth]{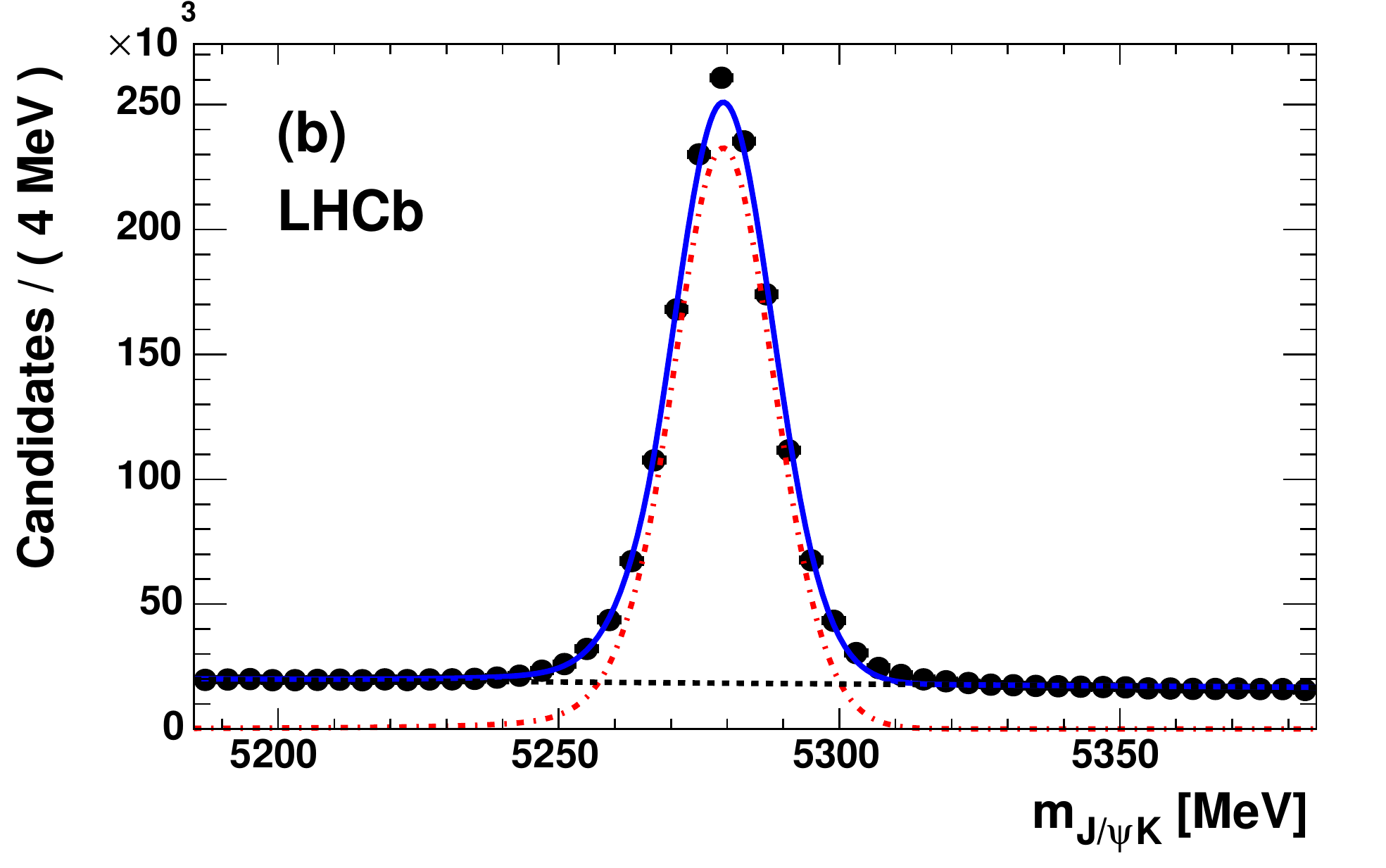}\\ \vspace{0.5cm}
\includegraphics[width=.45\textwidth]{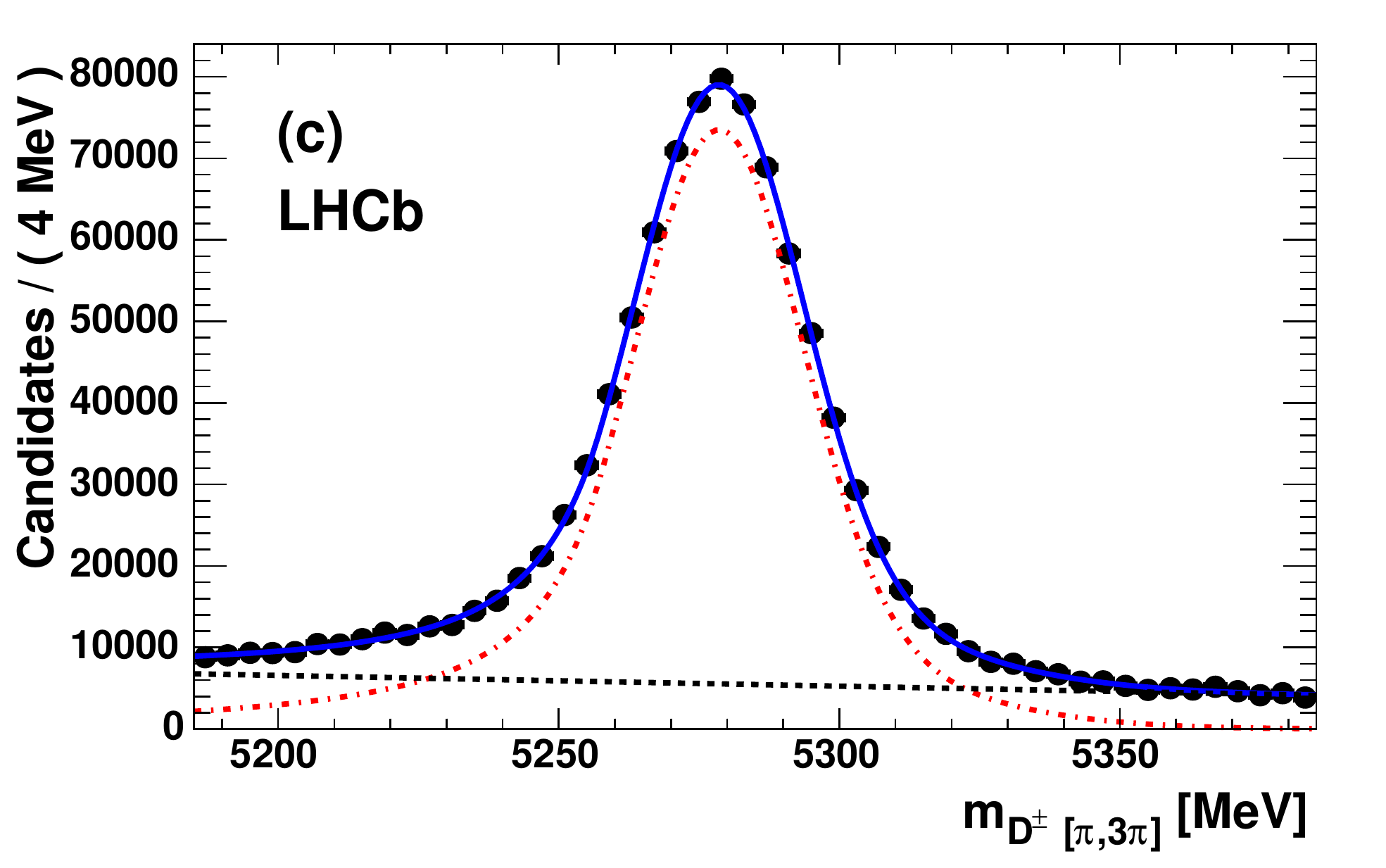}
\includegraphics[width=.45\textwidth]{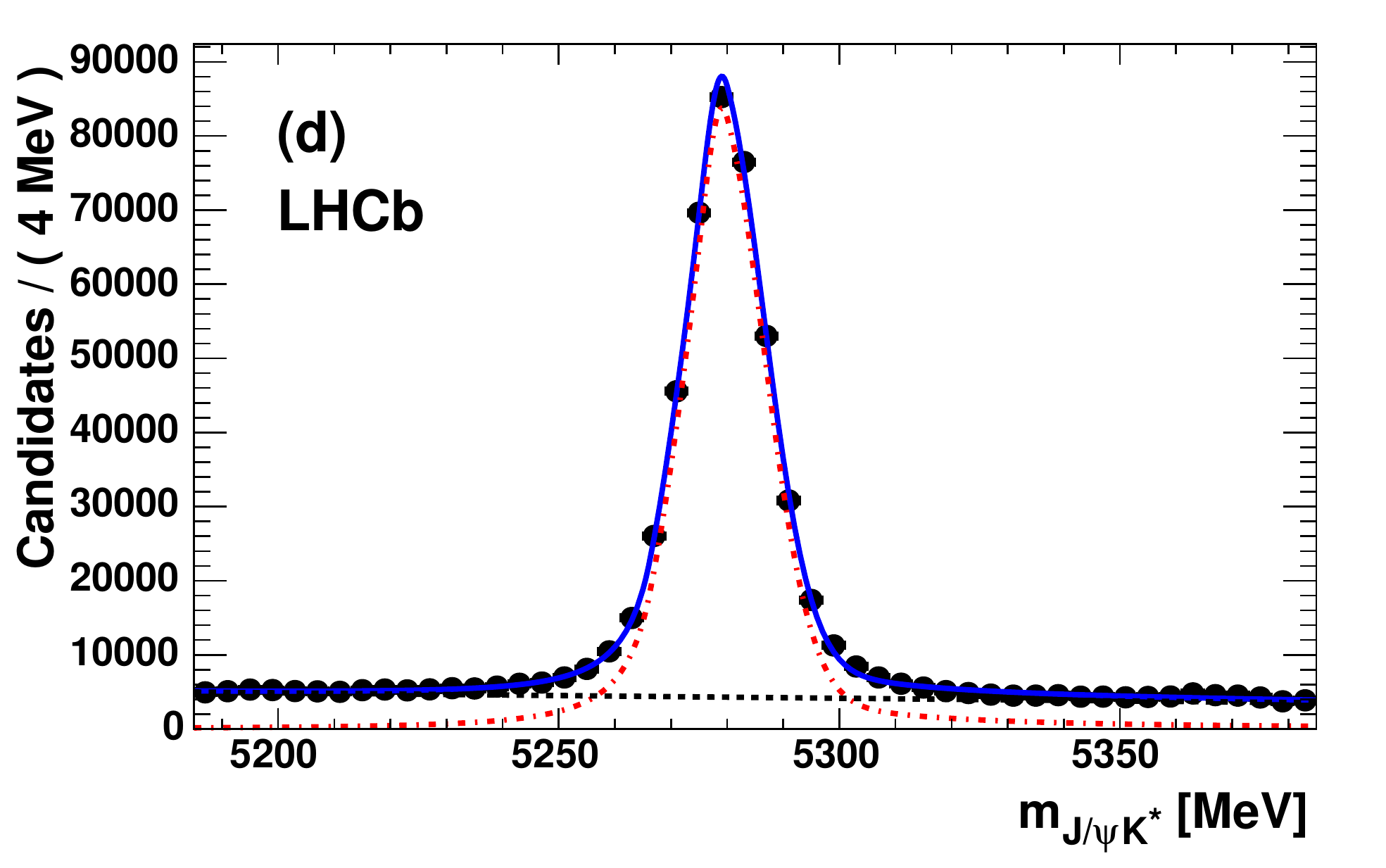}
\caption{
  Mass distributions of the \Bu and \Bz candidates reconstructed through
  (a) $\Bu\to \Dzb (\pi^+,\pi^+\pi^+\pi^-)$,
  (b) $\Bu\to J/\psi K^+$,
  (c) $\Bz\to \Dm (\pi^+,\pi^+\pi^+\pi^-)$, and
  (d) $\Bz\to J/\psi \Kstarz$ decays.
  The \jpsi, \Dzb and \Dm masses are constrained to their world average values~\cite{PDG2014}.
  Results of fits are superimposed for illustration.
  The signal (dot-dashed red line) is modelled with a double Crystal Ball~\cite{Skwarnicki:1986xj} distribution, while the background (dashed black line) is modelled with\
 a second-order polynomial.
  The total fit is shown as a solid blue line.
}
 \label{fig:bsignals}
\end{figure*}

The mass distributions for the \Bu and \Bz candidates are shown in Fig.~\ref{fig:bsignals}. 
Only \B meson candidates falling within $25 \mev$ of the nominal \B mass for the decay modes containing $\jpsi$ mesons, or $50 \mev$ for the other modes, are selected for further analysis. 
Samples of about 1.2 million \Bz and 2.5 million \Bu
candidates are obtained, with purity depending on decay mode and always larger than 80\%.
Each candidate is combined with any track that originates from the same PV and that is identified as a pion. 
The particle identification requirements on the companion pion are chosen to reduce potential backgrounds from misidentified particles to a level where they can be neglected in the analysis. 
Over the momentum range relevant for this analysis, the pion identification requirements are $81\%$ efficient at identifying pions, while they have  $3.1\%$ and $2.6\%$ probabilities respectively to misidentify a kaon or a proton as a pion. 
Since the production of \Bsss mesons is likely to be suppressed relative to the production of \Bss states, as has been observed for the ground states~\cite{LHCb-PAPER-2011-018,LHCb-PAPER-2012-037}, these requirements are expected to reduce background from the decays $B_{s1}(5830)^0 \to \Bus\Km$ and $B_{s2}^*(5840)^0 \to \Bus\Km \ {\rm or} \ \Bu\Km$, where the kaon is misidentified as a pion, to a negligible level.

Further selection requirements are placed on the \Bss candidate. 
The invariant mass and \chisqndf (ndf is the number of degrees of freedom) of the \Bss candidate vertex fit are calculated constraining the \B candidates and companion pion to originate from the
PV,
and also constraining the known \B meson mass, and the masses of intermediate \jpsi, \Dzb and \Dm mesons in the \B decay.
The \chisqndf of the \Bss candidate vertex fit is then required to be below $3.5$.
In order to reduce combinatorial backgrounds, the PV associated with the \Bss candidate is required to have fewer than 75 charged particles associated with it. 
The angle $\theta$ is required to satisfy $\cos \theta >-0.5$, where $\theta$ is the angle between the pion in the $\B\pi$ rest frame and the opposite direction of the boost vector from the $\B\pi$ rest frame to the laboratory frame.

Finally, the companion pion is required to have more than ($0.5$) $5\gev$ of (transverse) momentum, while the \B candidate is required to have $\pt > 10\gev$ for candidates where the companion pion has $\pt > 2\gev$.
In any selected event, the \B candidate can potentially be combined with several different pions to create \Bss candidates.
The average number of candidates per selected event is 1.4 and all of them are used for the subsequent analysis.

\section{Fit model}
\label{sec:Fitmodel}

\begin{figure}[!tb]
  \centering
  \includegraphics[width=0.99\textwidth]{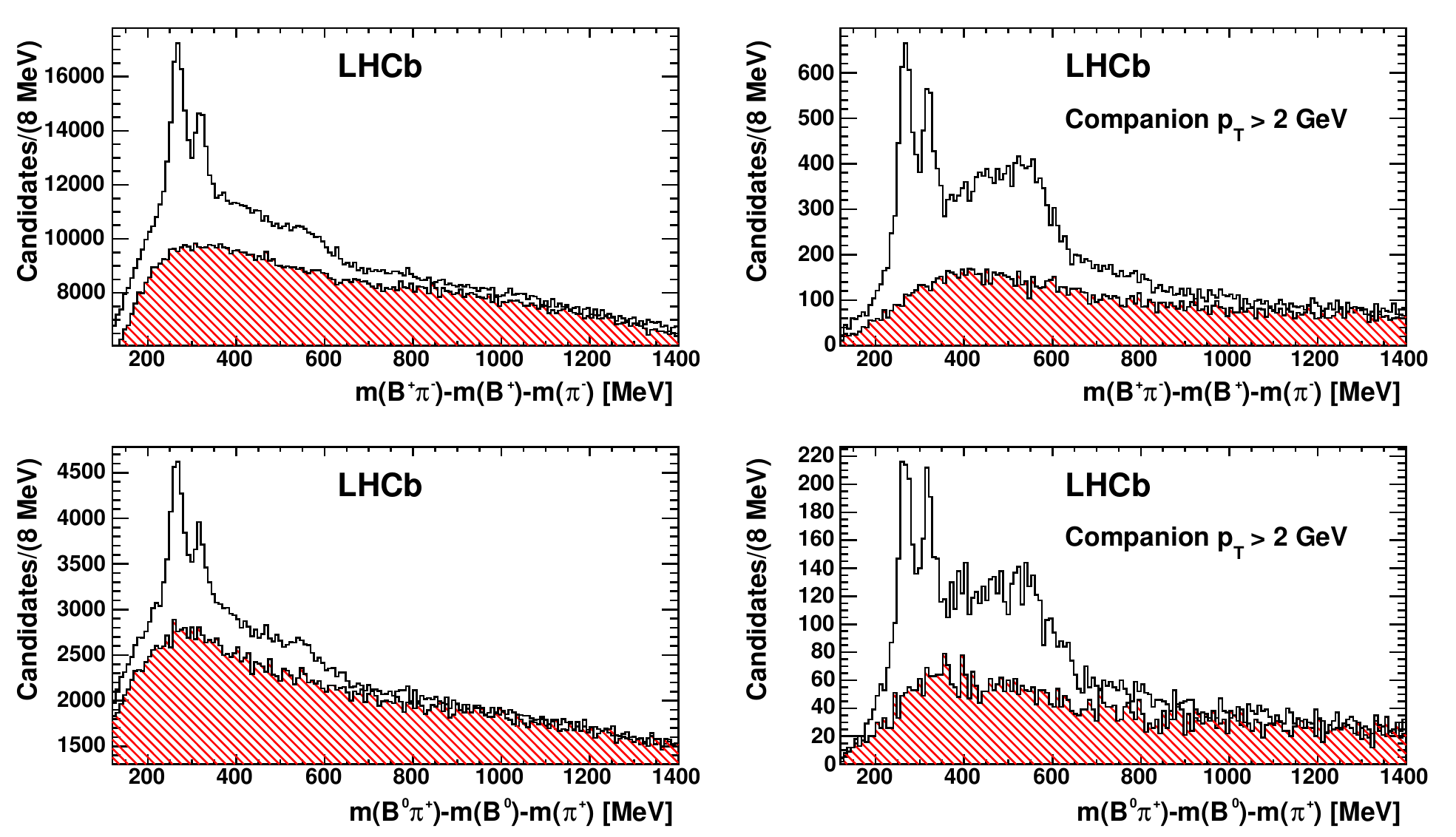}
  \caption{
    Distributions of the $Q$ values of the \Bss candidates after the selection for the (top) $\Bp$ and (bottom) $\Bz$ candidates. 
 The white histograms represent the RS combinations, while the overlaid shaded red histograms represent the WS combinations.
The
 right hand plots are made after applying an additional requirement of $\pt>2 \gev$ on
 the companion pion.
 \label{fig:bssrswsdistspt}
}
\end{figure}

The distributions of the mass difference, $Q \equiv m(\B\pi) - m(\B) - m(\pi)$, following these selection requirements are shown in Fig.~\ref{fig:bssrswsdistspt} for both RS and WS \Bss candidates, where $m_\B$ and $m_\pi$ are the known masses of the $B$ meson and the pion~\cite{PDG2014}.
All \B decay modes are combined in Fig.~\ref{fig:bssrswsdistspt} and in the subsequent analysis.
Two narrow peaks are seen in both $\Bp\pim$ and $\Bz\pip$ mass difference distributions, corresponding to the $B_1(5721)^{0,+}\to\Bstar\pi$ signal overlapping with the $B_2^*(5747)^{0,+}\to\Bstar\pi$ decay, and the $B_2^*(5747)^{0,+}\to\B\pi$ decay.
In addition, an excess of RS over WS combinations around $Q \sim 500\mev$ is particularly
prominent after requiring the companion pion to have $\pt > 2\gev$.
This peak could result from a combination of two heavier \Bss resonances, consistent with the expectation that \Bss states come in doublets, as described in Sec.~\ref{sec:Introduction}; the structure is further analysed as described below.
Furthermore, a comparison with the WS distributions shows a very broad excess of RS combinations lying under the resonances, corresponding to AP as discussed in Sec.~\ref{sec:Introduction}.

The $Q$-value distributions of $\Bp\pim$ and $\Bz\pip$ candidates are fitted independently to determine the masses and widths of the various resonant signals.
In order to increase sensitivity to the parameters of the high mass states,
the fits are performed in three bins of companion pion \pt: $0.5 < \pt \le 1
\gev$, $1 < \pt \le 2 \gev$ and $\pt > 2 \gev$.
The fits minimise the total $\chi^2$ of the $Q$-value distributions (in bins of width $1
\mev$) simultaneously for the three companion pion \pt bins.

The combinatorial background shape is obtained from WS combinations.
It has been checked that the WS background consists of purely combinatorial background by studying $\B\pi$ combinations in which a \B meson from one event is combined with a companion pion from another event; consistent shapes are found.   
The WS $Q$-value distributions are fitted with piecewise-defined, smooth polynomial (``spline'') functions.
The shape is fixed in the subsequent fit to the RS distribution, but the yield is allowed to vary.

Resonances are modelled with RBW lineshapes~\cite{Jackson:1964zd}, given by 
\begin{equation}
  \label{eq:bw}
  A_{\rm RBW}(m) = \frac{\Gamma(m)}{\left(m^2-m_0^2\right)^2+m_0^2\Gamma^2(m)}\,,
\end{equation}
where $m$ is the $\B\pion$ invariant mass (which is trivially related to the $Q$ value), $m_0$ is the mass value for the resonance\footnote{The mass difference $m_0 -m(B)-m(\pi)$ is referred to as the mean $\mu$ hereafter.} and $\Gamma(m)$ is the mass dependent width
\begin{equation}
  \Gamma(m) = \Gamma_0\frac{m_0}{m}\left(\frac{q(m)}{q(m_0)}\right)^{2l+1}{\mathcal F}_l^2\,.
\end{equation}
In the latter equation $\Gamma_0$ is the natural width, $q(m)$ is the \B or \pion momentum in the rest frame of the resonance and $l$ is the orbital angular momentum between the \B and \pion mesons.
The Blatt-Weisskopf form factors ${\mathcal F}_l$~\cite{Blatt-Weisskopf:1952,VonHippel:1972fg} account for the fact that the maximum angular momentum is limited by the phase-space in the decay. 
Defining the dimensionless quantity $z(m) = q^2(m)R^2$, where $R$ is the effective radius, ${\mathcal F_l}$ is defined as
\begin{eqnarray}
{\mathcal F_0} &=& 1\,, \cr
{\mathcal F_1} &=& \sqrt{\frac{1+z(m_0)}{1+z(m)}}\,,\cr
{\mathcal F_2} &=& \sqrt{\frac{\left(z(m_0)-3\right)^2+9z(m_0)}{\left(z(m)-3\right)^2+9z(m)}}\,.
\end{eqnarray}

Depending on the fit model, the \Bss resonances are described by five or six RBW shapes:
\begin{itemize}
\item{} one for the $B_1(5721)^{0,+}\to\Bstar\pi$ feed-down into the left narrow peak with width, yield, and mean  free to vary in the fits;
\item{} one for the $B_2^*(5747)^{0,+}\to\B\pi$ signal (the right narrow peak) with width, yield, and mean free to vary in the fits;
\item{} one for the $B_2^*(5747)^{0,+}\to\Bstar\pi$ feed-down into the left narrow peak with width fixed to be the same as that of the $B_2^*(5747)^{0,+}\to\B\pi$ signal, mean shifted from the $B_2^*(5747)^{0,+}\to\B\pi$ peak by the known $\Bstar-\B$ mass difference, $45.0\pm0.4$~\mev~\cite{PDG2014}, and relative yield in \pt bins constrained as described later;
\item{} two (or three) for the higher mass components, with widths, means, and yields free to vary in the fits (except in the three RBW case, where two of the means are constrained by the $\Bstar-\B$ mass difference). 
\end{itemize}
The alternative descriptions for the higher mass resonances are motivated by the lack of knowledge of their quantum numbers.
As described in Sec.~\ref{sec:Introduction}, a doublet of states is expected to give rise to three peaks.
For example, for the $(B(2{\rm S}),B^*(2{\rm S}))$ doublet the higher (lower) mass of the pair has natural (unnatural) spin-parity.
The description with three RBW shapes, two of which are constrained to have means offset by the $\Bstar-\B$ mass difference, is therefore a
physically motivated choice, obtained by applying quark-model expectations to the new states. 
However, there are two possibilities for this configuration, since it may be either the lower or the higher of the states that gives rise to two peaks.
The alternative, with only two RBW shapes, is an empirical model, that corresponds to the minimal choice necessary to obtain a satisfactory description of the data.
This is taken as the default and is referred to hereafter as the empirical model, but results of alternative fits with three RBW shapes are also presented.

The RBW shapes have several parameters which need to be fixed in the fits, in particular the spin and effective radius input to the Blatt-Weisskopf form factors.
The $B_1(5721)^{0,+}$ and $B_2^*(5747)^{0,+}$ resonances are assigned spin 1 and 2, respectively, and are both assumed to decay via D-wave ($l = 2$), while
the two higher mass resonances are assigned spin 0 ($l=0$) in the default fit.
The effective radius is fixed to $4\gev^{-1}$~\cite{LHCb-PAPER-2014-036}.
The mass resolution is around $2\mev$ which is negligible compared to the natural widths ($> 20\mev$) of the resonances, and is therefore not modelled.
The variation of the signal reconstruction efficiency with $Q$ value is described with a fifth-order polynomial function with parameters determined from simulation.
All signal parameters except the yields are shared between the different \pt bins and \B meson decay modes,
though the efficiency function is determined independently for each \pt bin.

The AP component is caused by correlations between the \B meson and the companion pion, and as such is not present in either the WS sample or in a sample obtained by mixing \B mesons and pions from different events. 
As there is no suitable data control sample from which it can be constrained, it must be empirically modelled. 
The AP is modelled by a sixth-order polynomial shape determined from simulation with an additional broad spin-0 RBW function to account for possible data-simulation differences.
The latter component is introduced since the modelling of fragmentation effects in the simulation is expected to be imprecise.

The relative yields of $B_2^*(5747)^{0,+}\to\Bstar\pi$ and $\B\pi$ in each \pt bin are fixed according to the relative efficiencies found in simulation, so that the relative branching fraction ratios ${\cal B}(B_2^*(5747)^{0,+} \to B^* \pi)/{\cal B}(B_2^*(5747)^{0,+} \to B \pi)$ are free parameters of the fits.
The WS and AP 
yields are freely varied in the fits, independently in each \pt bin. The RBW parameters of the AP shape are also
freely varied; the remaining parameters are fixed to the values obtained from simulation to avoid instabilities in the fits.
The fit procedure is validated using large ensembles of pseudoexperiments.

\section{Fit results}
\label{sec:nominalfit}
\begin{figure}[!thb]
  \centering
  \includegraphics[width=0.9\textwidth]{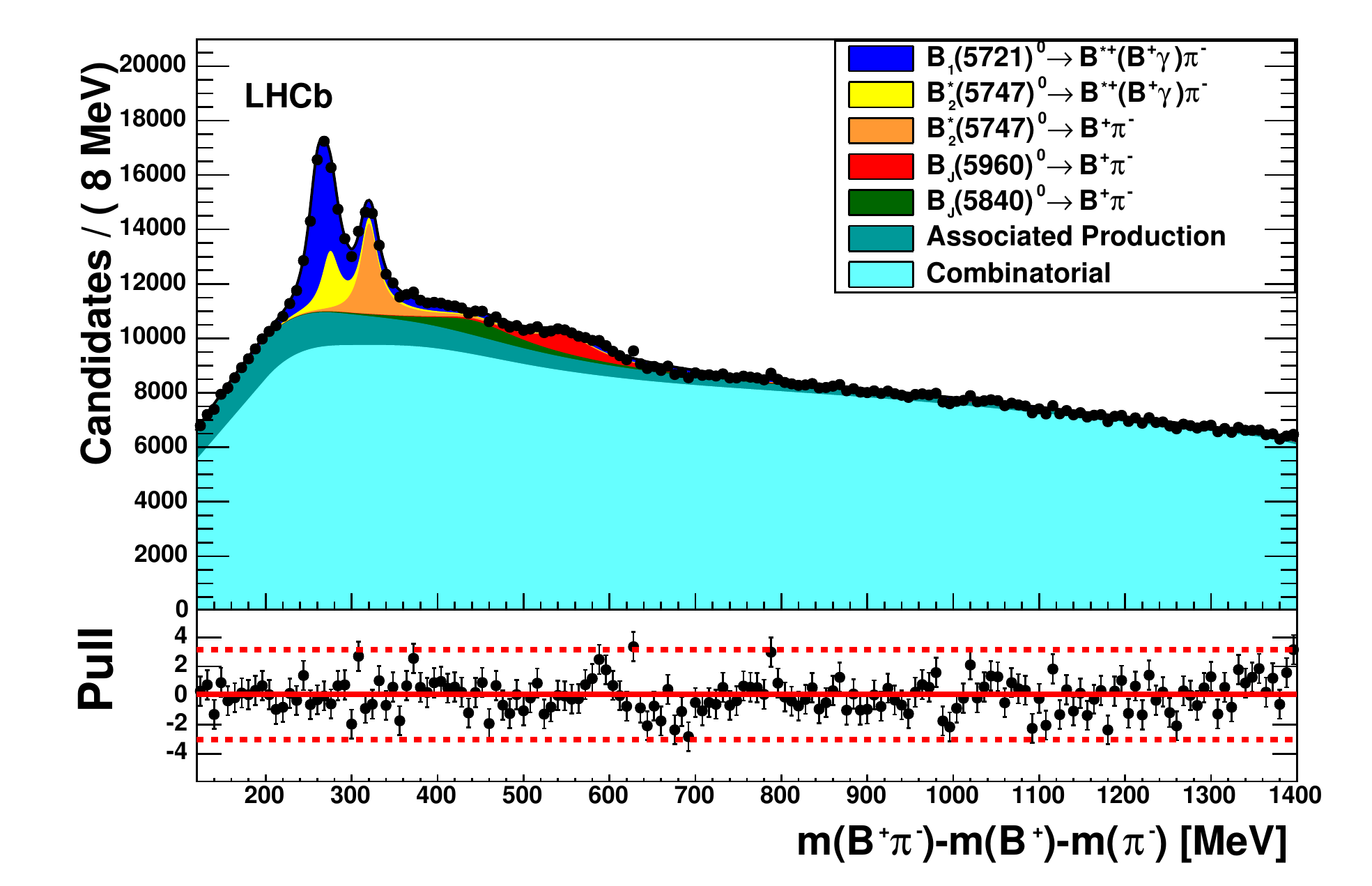}
  \includegraphics[width=0.9\textwidth]{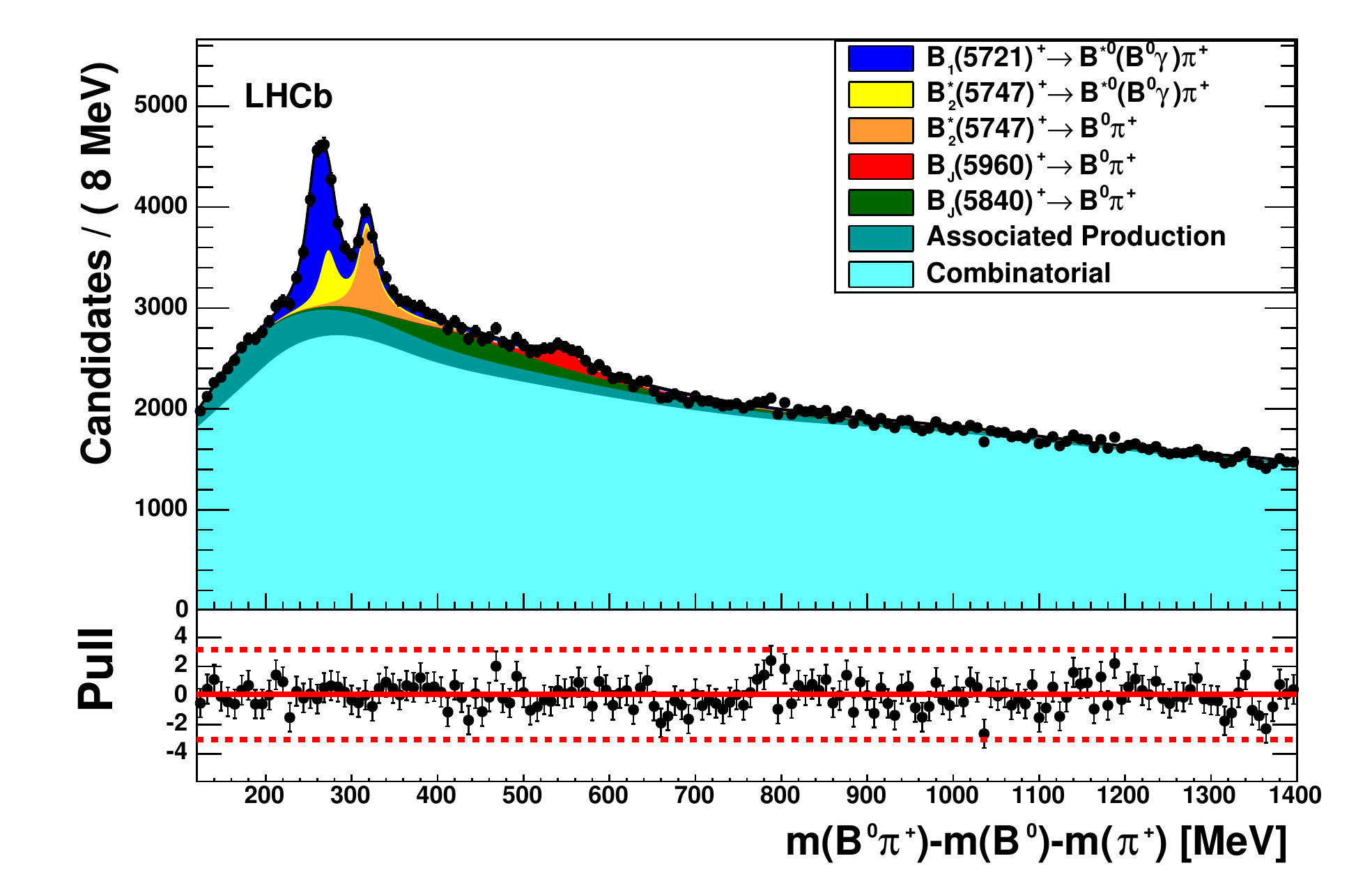}
  \caption{
    Result of the fits to the $Q$-value distributions for (top) $\Bp\pim$ and (bottom) $\Bz\pip$ candidates.
    The components are labelled in the legend. 
    The normalised residuals (pulls) of the difference between the fit results and the data points, 
    divided by their uncertainties, are shown underneath each plot.
  }
  \label{fig:nominalfit}
\end{figure}

\begin{figure}[!thb]
  \centering
  \includegraphics[width=0.496\textwidth]{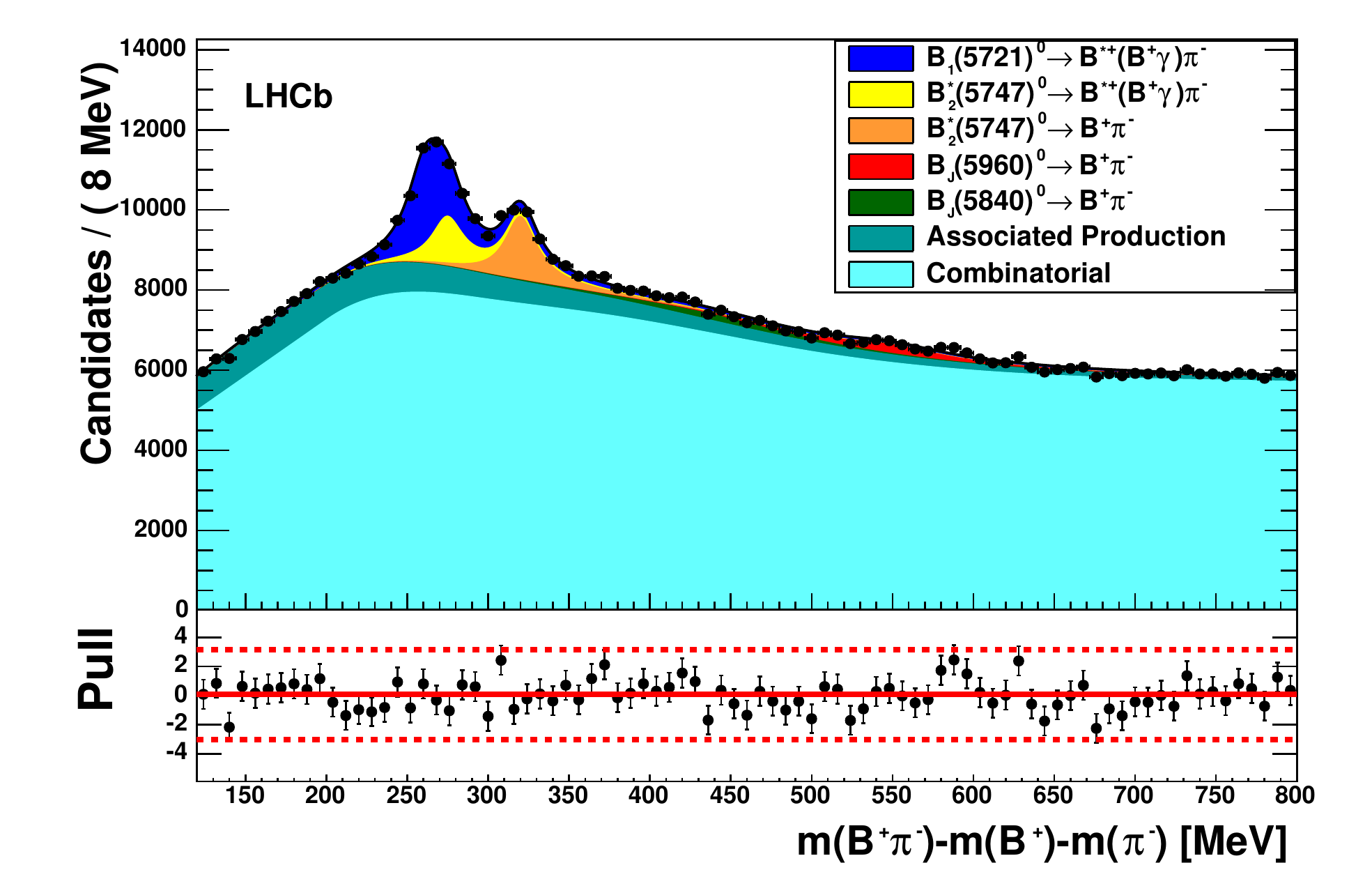}
  \put(-155,128){\textbf{\tiny{$1 \geq \pt > 0.5$ GeV}}}
  \includegraphics[width=0.496\textwidth]{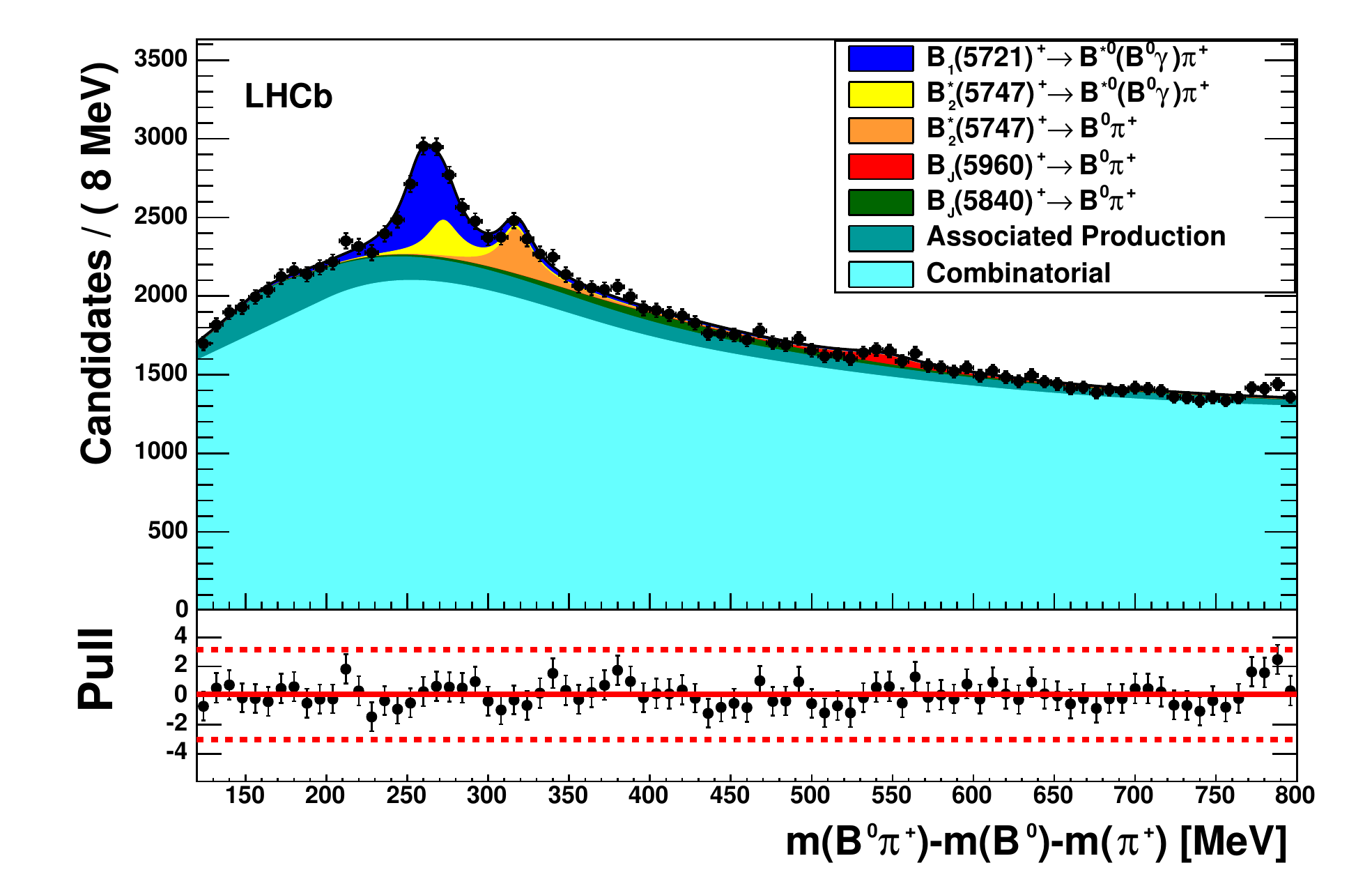}
  \put(-155,128){\textbf{\tiny{$1 \geq \pt > 0.5$ GeV}}}
  \\
  \includegraphics[width=0.496\textwidth]{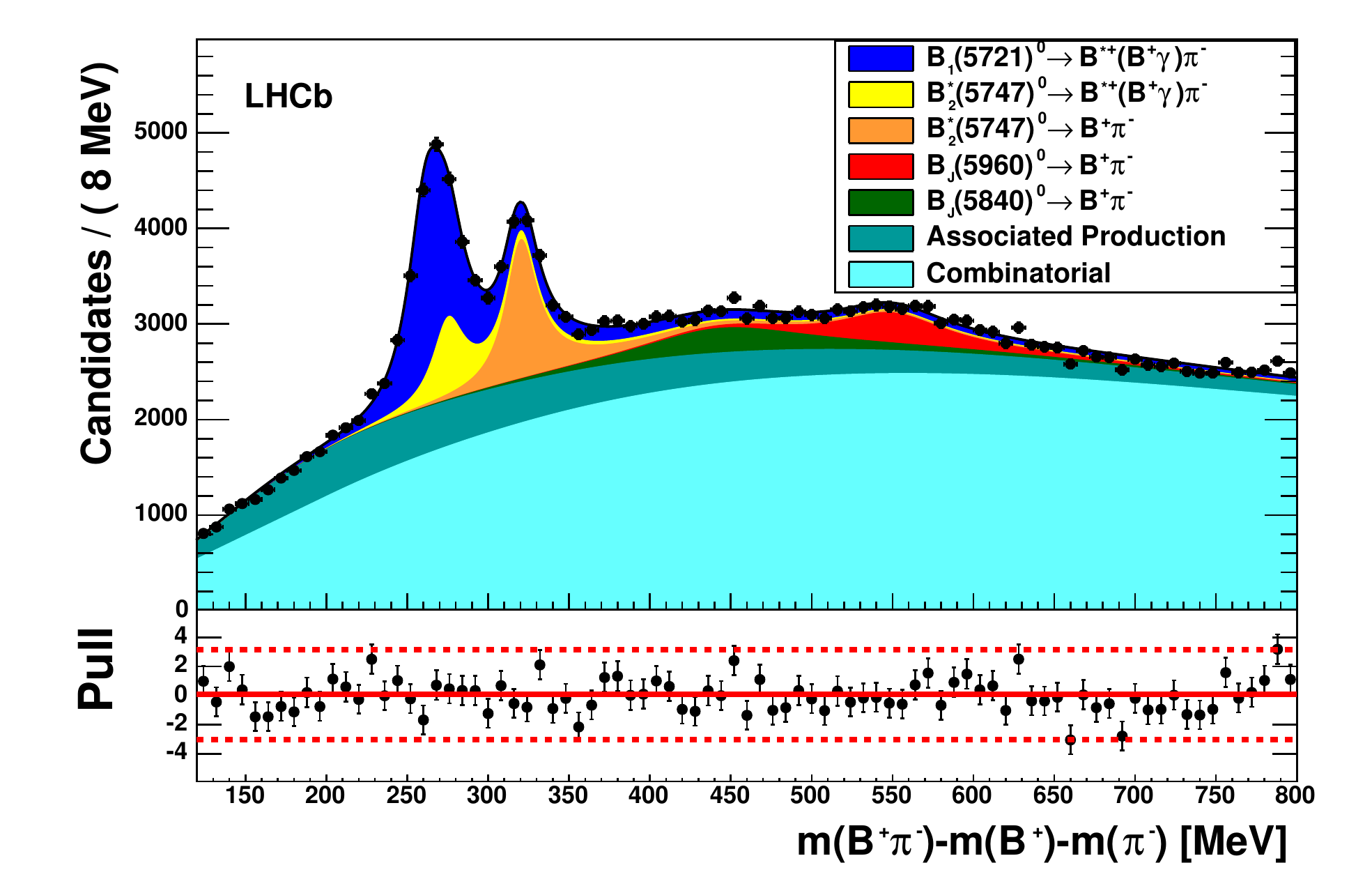}
  \put(-155,128){\textbf{\tiny{$2 \geq \pt > 1$ GeV}}}
  \includegraphics[width=0.496\textwidth]{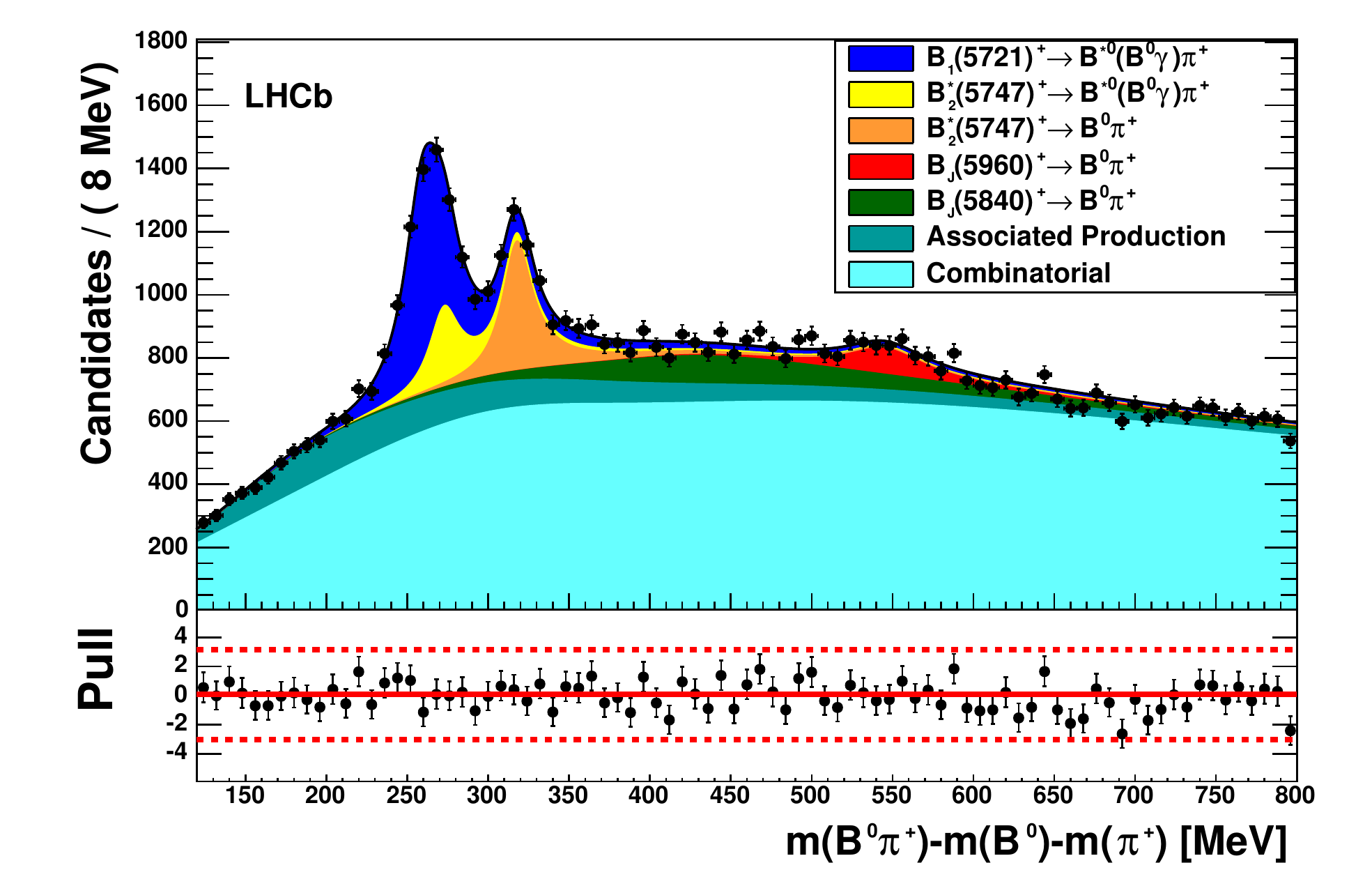}
  \put(-155,128){\textbf{\tiny{$2 \geq \pt > 1$ GeV}}}
  \\
  \includegraphics[width=0.496\textwidth]{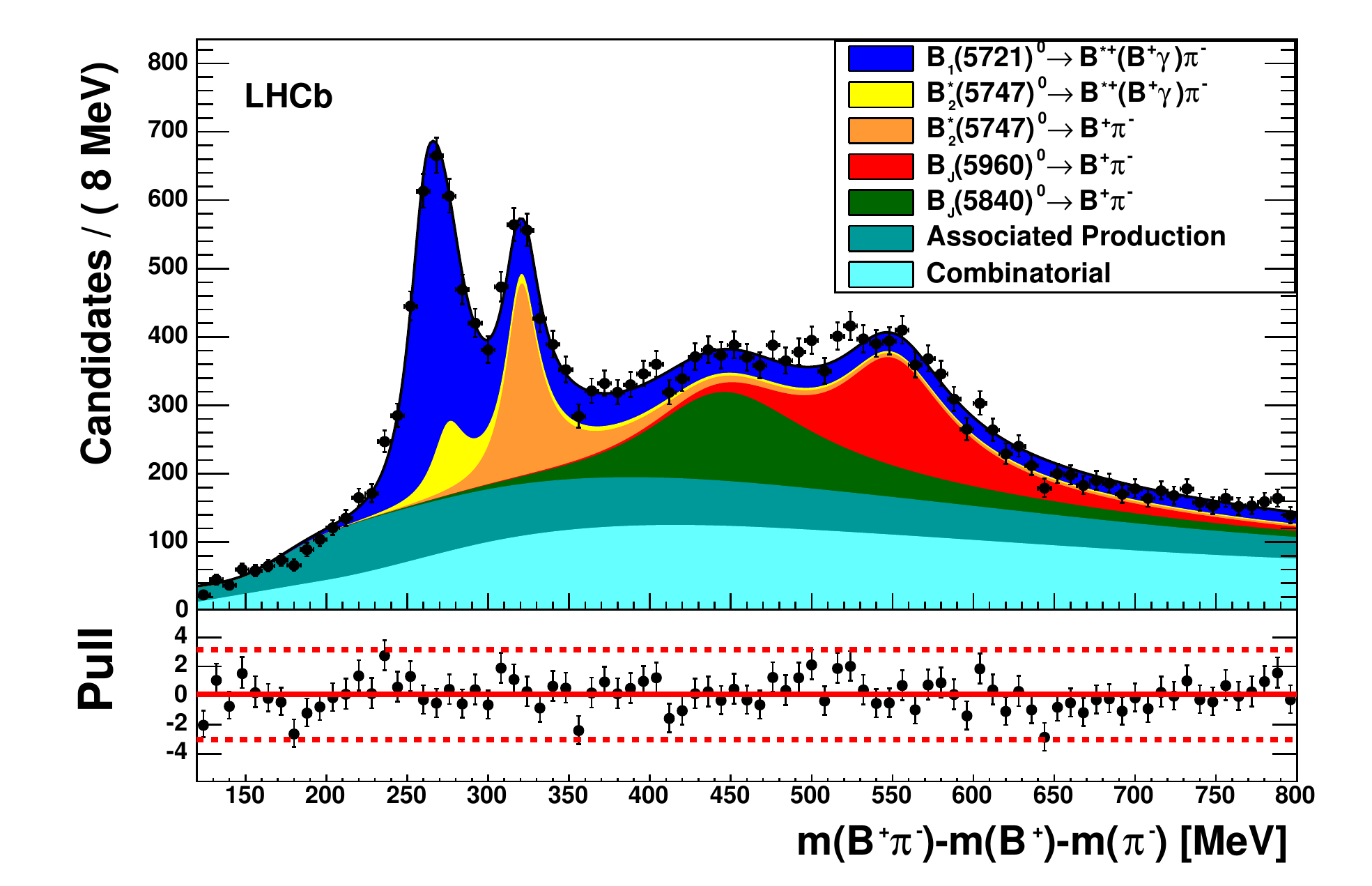}
  \put(-155,128){\textbf{\tiny{$\pt > 2$ GeV}}}
  \includegraphics[width=0.496\textwidth]{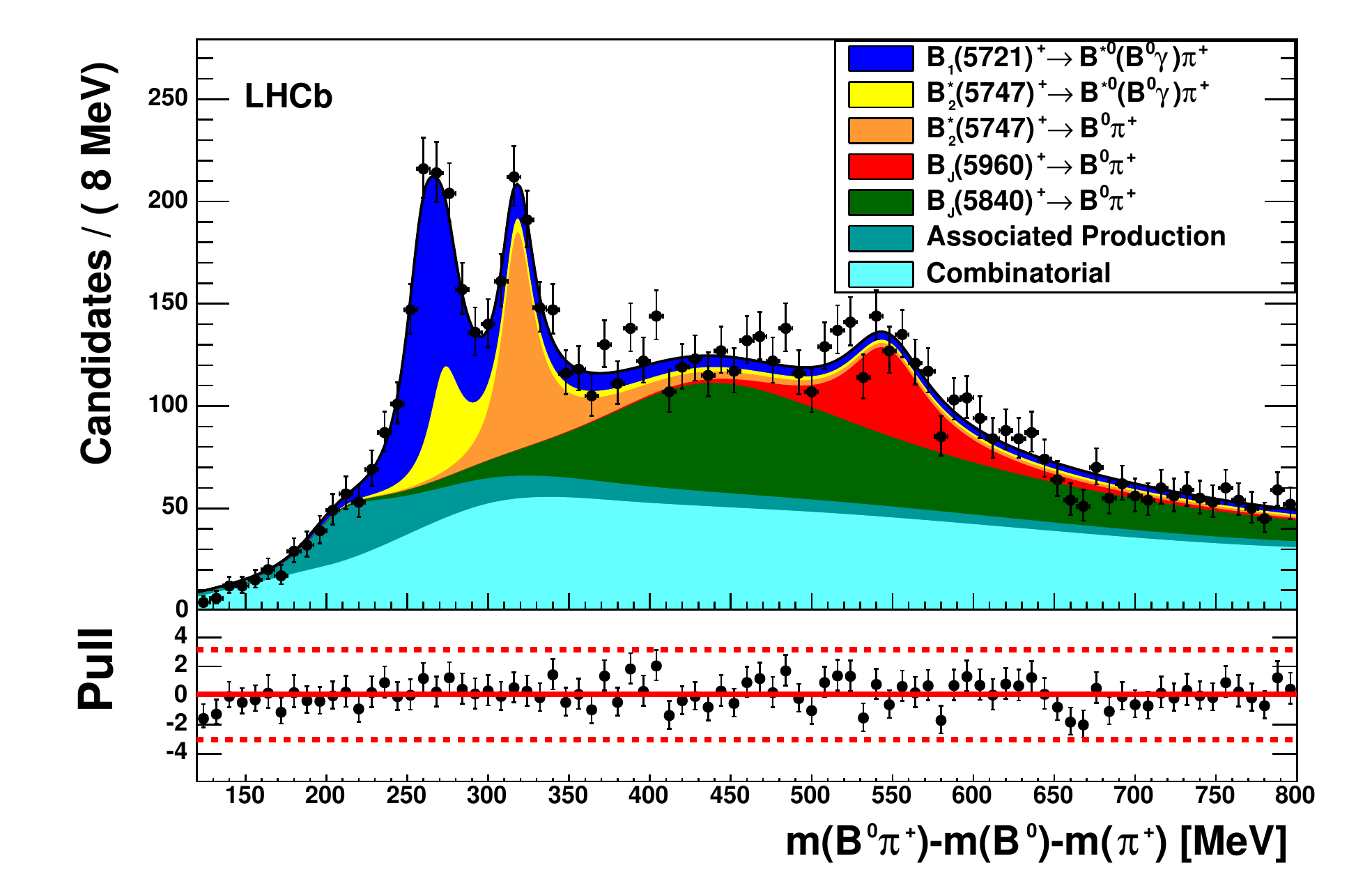}
  \put(-155,128){\textbf{\tiny{$\pt > 2$ GeV}}}
  \caption{
    Result of the fit to (left) $\Bp\pim$ and (right) $\Bz\pip$ candidates, split into (top to bottom) low, medium and high \pt bins,
    with ranges as labelled on the plots. The components are labelled in the legends. The fit pulls are shown underneath each plot.
  }
  \label{fig:nominalfit_rsbp_byPT}
\end{figure}

The results of the empirical model fits to the \Bss candidates integrated over the three \pt bins are shown in Fig.~\ref{fig:nominalfit}.
The results are also shown split by \pt bin in Fig.~\ref{fig:nominalfit_rsbp_byPT}, where the plots have been zoomed into the range below $800 \mev$ in order
to emphasise the resonant structures.
The results for the parameters of interest are reported in Table~\ref{tab:nominalfitresults}.
\begin{table}[!b]
  \centering
  \caption{
    Results of the fits when two RBW functions are used for the $B_J(5840)^{0,+}$ and $B_J(5960)^{0,+}$ states (empirical model).
    The mean $\mu$ of each peak is given together with the width $\Gamma$ and the yield $N_{\textrm{state}}$.
    The parameters related to the AP and WS components are suppressed for brevity.
    All uncertainties are statistical only.
    Units of \mev for $\mu$ and $\Gamma$ are implied.
  }
  \label{tab:nominalfitresults}
  \vspace{-1ex}
  \begin{tabular}{lcc}
    \hline
    Fit parameter & $\phanii\Bp\pim$ & $\phanii\Bz\pip$\\
    \hline
    $B_1(5721)^{0,+}$ $\mu$      & $263.9 \pm 0.7$ & $260.9 \pm 1.8$ \\
    $B_1(5721)^{0,+}$ $\Gamma$   & $\phani30.1 \pm 1.5$ & $\phani29.1 \pm 3.6$ \\
    $B_2^*(5747)^{0,+}$ $\mu$    & $320.6 \pm 0.4$ & $318.1 \pm 0.7$ \\
    $B_2^*(5747)^{0,+}$ $\Gamma$ & $\phani24.5 \pm 1.0$ & $\phani23.6 \pm 2.0$ \\
    \hline
    $N_{B_1(5721)^{0,+}}$ low \pt   &  $14200 \pm 1400$  & $3140 \pm 750$ \\
    $N_{B_1(5721)^{0,+}}$ mid \pt   &  $16200 \pm 1500$  & $4020 \pm 890$ \\
    $N_{B_1(5721)^{0,+}}$ high \pt  &  $\phani4830 \pm \phani470$  & $\phani940 \pm 260$ \\
    $N_{B_2^*(5747)^{0,+}}$ low \pt &  $\phani7450 \pm \phani420$  & $1310 \pm 180$ \\
    $N_{B_2^*(5747)^{0,+}}$ mid \pt &  $\phani7600 \pm \phani340$  & $2070 \pm 180$ \\
    $N_{B_2^*(5747)^{0,+}}$ high \pt&  $\phani1690 \pm \phani130$  & $\phani640 \pm \phani80$ \\
${\cal B}(B_2^*(5747)^{0,+} \to B^* \pi)/{\cal B}(B_2^*(5747)^{0,+} \to B \pi)$ & $\phani0.71 \pm 0.14$ & $\phani1.0 \pm 0.5$ \\
    \hline
    $B_J(5840)^{0,+}$ $\mu$    & $444 \pm \phanid5$      & $431 \pm \phand13$ \\
    $B_J(5840)^{0,+}$ $\Gamma$ & $127 \pm \phand17$      & $224 \pm \phand24$ \\
    $B_J(5960)^{0,+}$ $\mu$    & $550.4 \pm 2.9\phanid$  & $545.8 \pm 4.1\phanid$ \\
    $B_J(5960)^{0,+}$ $\Gamma$ & $\phani82 \pm \phanid8$ & $\phani63 \pm \phand15$ \\
    \hline
    $N_{B_J(5840)^{0,+}}$ low \pt  & $3200 \pm 1300$     & $1630 \pm 970$ \\
    $N_{B_J(5840)^{0,+}}$ mid \pt  & $5600 \pm 1000$     & $3230 \pm 720$ \\
    $N_{B_J(5840)^{0,+}}$ high \pt & $3090 \pm \phani550$     & $2280 \pm 450$ \\
    $N_{B_J(5960)^{0,+}}$ low \pt  & $3270 \pm \phani660$     & $\phani610 \pm 240$ \\
    $N_{B_J(5960)^{0,+}}$ mid \pt  & $4590 \pm \phani610$     & $\phani910 \pm 250$ \\
    $N_{B_J(5960)^{0,+}}$ high \pt & $2400 \pm \phani320$     & $\phani500 \pm 140$ \\
    \hline
  \end{tabular}
\end{table}
Note that the reported mean values correspond to the peak positions, and do not include any correction for the $\Bstar-\B$ mass difference, but when a state is assumed to have natural spin-parity, and therefore gives two peaks, the mass value reported is that of the higher peak. The results are consistent for the charged and neutral states, as expected since the uncertainties are larger than isospin  splitting effects.
The results for the higher mass states depend on whether they are assumed to have natural or unnatural spin-parity, and the results with the alternative hypotheses are presented in Table~\ref{tab:naturalJPfitresults}. For the purpose of labelling, and without prejudice on their quantum numbers, the lower of these states is referred to subsequently as the $B_J(5840)^{0,+}$ and the other as the $B_J(5960)^{0,+}$ state.

\clearpage

The covariance matrix of the empirical model fit is given in Appendix~\ref{app:cov_syst_err}. 
For brevity, the results for the signal yields and for the background parameters are not reported. The magnitudes of the correlations between the signal observables and background shapes are smaller than $30\%$.
All fits have acceptable minimum \chisq values.
\begin{table}[!tb]
  \centering
  \caption{
    Results of the fits when the natural spin-parity hypothesis is assigned to (top quadruplet) the $B_J(5840)^{0,+}$ state or (bottom quadruplet) the $B_J(5960)^{0,+}$ state, so that three RBW shapes are used to model the broad resonances in the fit.
    The mean $\mu$ of each peak is given together with the width $\Gamma$.
    All uncertainties are statistical only. 
    Units of \mev for $\mu$ and $\Gamma$ are implied.
  }
  \label{tab:naturalJPfitresults}  
  \vspace{-1ex}
  \begin{tabular}{lcc}
    \hline
    Fit parameter & $\Bp\pim$ & $\Bz\pip$ \\
    \hline
    $B_J(5840)^{0,+}$ $\mu$    & $471 \pm 22$      & $455 \pm 26$  \\
    $B_J(5840)^{0,+}$ $\Gamma$ & $107 \pm 20$      & $215 \pm 27$  \\
    $B_J(5960)^{0,+}$ $\mu$    & $552 \pm \phani4$ & $547 \pm \phani4$  \\
    $B_J(5960)^{0,+}$ $\Gamma$ & $\phani82 \pm 10$      & $\phani61 \pm 15$  \\  
    \hline
    $B_J(5840)^{0,+}$ $\mu$    & $444 \pm \phani5$      & $425 \pm 15$ \\ 
    $B_J(5840)^{0,+}$ $\Gamma$ & $119 \pm 17$           & $229 \pm 27$ \\  
    $B_J(5960)^{0,+}$ $\mu$    & $575 \pm \phani6$      & $547 \pm \phani5$  \\ 
    $B_J(5960)^{0,+}$ $\Gamma$ & $\phani56 \pm \phani7$ & $\phani61 \pm 14$  \\   
    \hline
  \end{tabular}
\vspace{-2ex}
\end{table}

\section{Systematic uncertainties}
\label{sec:systematics}
Systematic uncertainties are evaluated in a data-driven manner, by varying fit parameters or configurations from their default values and taking the difference in the fit results as a systematic uncertainty. 
Summaries of the systematic uncertainties are given in Tables~\ref{tab:totalsysttable_bp} and~\ref{tab:totalsysttable_bz} for $\Bp\pim$ and $\Bz\pip$ resonances.
 The total systematic uncertainties on each individual parameter are obtained by combining all sources in quadrature.
\begin{table}[!tb]
  \centering
  \caption{
    Systematic uncertainties on the results of the fit to the $\Bp\pim$ candidates.
    Units of \mev for $\mu$ and $\Gamma$ are implied.
  }
  \label{tab:totalsysttable_bp}
  \vspace{-1ex}
  \resizebox{\textwidth}{!}{
  \begin{tabular}{lccccccccc}
    \hline
    Source & \multicolumn{2}{c}{$B_1(5721)^0$} & \multicolumn{3}{c}{$B_2^*(5747)^0$} & \multicolumn{2}{c}{$B_J(5840)^0$} & \multicolumn{2}{c}{$B_J(5960)^0$} \\
    & $\mu$ & $\Gamma$ & BF ratio & $\mu$ & $\Gamma$ & $\mu$ & $\Gamma$ & $\mu$ & $\Gamma$ \\
\hline
Total statistical & 0.72 & 1.52 & 0.14 & 0.37 & 1.01 & 4.95 & 16.70 & 2.88 & 7.71 \\
\hline
Fit range (high) & 0.33 & 1.30 & 0.06 & 0.08 & 0.37 & 2.20 & \phani2.90 & 0.52 & 0.26 \\
Fit range (low) & 0.04 & 0.11 & 0.01 & 0.02 & 0.39 & 0.04 & \phani8.22 & 0.69 & 2.83 \\
$2 \mev$ bins & 0.02 & 0.14 & 0.00 & 0.04 & 0.07 & 1.09 & \phani0.50 & 0.08 & 1.00 \\
Spline knots & 0.11 & 0.01 & 0.02 & 0.02 & 0.26 & 1.75 & \phani0.04 & 0.45 & 1.44 \\
Float AP & 0.03 & 0.00 & 0.00 & 0.03 & 0.30 & 1.58 & 10.16 & 0.73 & 4.23 \\
\hline
$B_2^*(5747)^0$ rel. eff., low \pt & 0.56 & 0.91 & 0.15 & 0.08 & 0.16 & 0.07 & \phani0.23 & 0.00 & 0.18 \\
$B_2^*(5747)^0$ rel. eff., mid \pt & 0.64 & 1.01 & 0.05 & 0.09 & 0.18 & 0.08 & \phani0.26 & 0.00 & 0.16 \\
$B_2^*(5747)^0$ rel. eff., high \pt & 0.20 & 0.37 & 0.03 & 0.02 & 0.07 & 0.02 & \phani0.00 & 0.01 & 0.09 \\
Eff. variation with $Q$ value & 0.13 & 0.33 & 0.02 & 0.04 & 0.07 & 0.45 & \phani2.46 & 0.19 & 0.70 \\
Data-simulation reweighting & 0.07 & 0.38 & 0.02 & 0.00 & 0.16 & 1.81 & \phani2.03 & 0.49 & 0.12 \\
\B \pt  & 0.02 & 0.20 & 0.01 & 0.24 & 0.72 & 3.98 & \phani3.67 & 1.30 & 4.29 \\
\pt binning & 0.90 & 2.45 & 0.24 & 0.06 & 0.39 & 1.49 & 27.77 & 4.20 & 1.47 \\
\hline
Fit bias & 0.06 & 0.17 & 0.01 & 0.00 & 0.16 & 0.45 & \phani5.34 & 0.40 & 2.24 \\
\hline
Spin & 0.02 & 0.06 & 0.01 & 0.06 & 0.46 & 1.95 & \phani3.32 & 0.62 & 3.74 \\
Effective radius & 0.33 & 1.44 & 0.02 & 0.12 & 0.76 & 2.17 & \phani9.68 & 1.24 & 3.81 \\
$\Bstar-\B$ mass & 0.10 & 0.11 & 0.03 & 0.02 & 0.11 & 0.04 & \phani0.17 & 0.00 & 0.09 \\
$B_J(5840)^0$ $J^P$ & 0.01 & 0.04 & 0.00 & 0.01 & 0.01 &  ---  &  ---  & 1.67 & 0.76 \\
$B_J(5960)^0$ $J^P$ & 0.01 & 0.20 & 0.00 & 0.00 & 0.16 & 0.18 & \phani8.00 &  ---  &  ---  \\
Extra state & 0.00 & 0.26 & 0.00 & 0.04 & 0.34 & 1.67 & \phani0.99 & 0.12 & 2.08 \\
\hline
Total systematic  & 1.36 & 3.49 & 0.30 & 0.33 & 1.48 & 6.68 & 34.24 & 5.10 & 9.41 \\
\hline
  \end{tabular}
}
\vspace{-2ex}
\end{table}
\begin{table}[!tb]
  \centering
  \caption{
    Systematic uncertainties on the results of the fit to the $\Bz\pip$ candidates.
    Units of \mev for $\mu$ and $\Gamma$ are implied.
  }
  \label{tab:totalsysttable_bz}
  \vspace{-1ex}
  \resizebox{\textwidth}{!}{
  \begin{tabular}{lccccccccc}
    \hline
    Source & \multicolumn{2}{c}{$B_1(5721)^+$} & \multicolumn{3}{c}{$B_2^*(5747)^+$} & \multicolumn{2}{c}{$B_J(5840)^+$} & \multicolumn{2}{c}{$B_J(5960)^+$} \\
    & $\mu$ & $\Gamma$ & BF ratio & $\mu$ & $\Gamma$ & $\mu$ & $\Gamma$ & $\mu$ & $\Gamma$ \\
\hline
Total statistical & 1.81 & 3.57 & 0.51 & 0.72 & 1.99 & 12.70 & 23.90 & 4.07 & 14.50 \\
\hline
Fit range (high) & 0.35 & 0.74 & 0.10 & 0.11 & 0.25 & \phani1.51 & 12.85 & 0.38 & \phani0.46 \\
Fit range (low) & 0.64 & 1.13 & 0.13 & 0.06 & 0.13 & \phani7.85 & 39.71 & 0.14 & \phani1.44 \\
$2 \mev$ bins & 0.16 & 0.34 & 0.05 & 0.10 & 0.49 & \phani0.58 & \phani3.84 & 0.28 & \phani0.52 \\
Spline knots & 0.30 & 0.08 & 0.07 & 0.03 & 0.22 & \phani1.94 & \phani2.64 & 0.25 & \phani0.25 \\
Float AP & 0.02 & 0.31 & 0.01 & 0.02 & 0.03 & \phani2.91 & \phani2.44 & 0.19 & \phani2.24 \\
\hline
$B_2^*(5747)^+$ rel. eff, low \pt & 1.50 & 2.14 & 0.43 & 0.12 & 0.49 & \phani0.15 & \phani1.63 & 0.02 & \phani0.03 \\
$B_2^*(5747)^+$ rel. eff, mid \pt & 1.55 & 2.26 & 0.53 & 0.12 & 0.51 & \phani0.29 & \phani2.03 & 0.04 & \phani0.15 \\
$B_2^*(5747)^+$ rel. eff, high \pt & 0.49 & 0.90 & 0.11 & 0.03 & 0.12 & \phani0.10 & \phani0.84 & 0.02 & \phani0.07 \\
Eff. variation with $Q$ value & 0.07 & 0.27 & 0.02 & 0.03 & 0.10 & \phani1.65 & \phani7.28 & 0.16 & \phani0.94 \\
Data-simulation reweighting & 0.04 & 0.38 & 0.03 & 0.00 & 0.02 & \phani2.13 & \phani7.49 & 0.40 & \phani1.75 \\
\B \pt  & 0.45 & 1.38 & 0.17 & 0.14 & 0.54 & \phani1.16 & \phani7.79 & 0.98 & \phani4.65 \\
\pt binning & 1.82 & 1.03 & 0.26 & 0.15 & 1.38 & \phani0.54 & 55.56 & 0.94 & 11.43 \\
\hline
Fit bias & 0.14 & 0.39 & 0.04 & 0.01 & 0.32 & \phani1.14 & \phani7.65 & 0.57 & \phani4.21 \\
\hline
Spin & 0.14 & 0.33 & 0.05 & 0.15 & 0.94 & \phani4.18 & 24.49 & 1.67 & \phani5.98 \\
Effective radius & 0.70 & 1.48 & 0.12 & 0.19 & 0.29 & \phani2.82 & 22.15 & 0.39 & \phani3.76 \\
$\Bstar-\B$ mass & 0.21 & 0.06 & 0.07 & 0.01 & 0.15 & \phani0.32 & \phani0.48 & 0.03 & \phani0.07 \\
$B_J(5840)^+$ $J^P$ & 0.00 & 0.05 & 0.00 & 0.03 & 0.15 &  ---  &  ---  & 0.72 & \phani1.64 \\
$B_J(5960)^+$ $J^P$ & 0.02 & 0.01 & 0.01 & 0.04 & 0.26 & \phani5.99 & \phani4.86 &  ---  &  ---  \\
Extra state & 0.03 & 0.41 & 0.00 & 0.00 & 0.15 & \phani6.28 & 12.82 & 0.43 & \phani7.81 \\
\hline
Total systematic & 3.10 & 4.28 & 0.79 & 0.40 & 2.07 & 13.70 & 79.82 & 2.52 & 17.18 \\
\hline
  \end{tabular}
}
\vspace{-2ex}
\end{table}
The covariance matrix of the systematic uncertainties, given in Appendix~\ref{app:cov_syst_err}, is computed by considering the correlated effects on the fit parameters of the systematic uncertainties.

The modelling of the background shapes may depend on the choice of fit range.  
The upper and lower edges of the range are varied independently by 20\% to assign systematic uncertainties.
Similarly, any dependence of the results on the choice of bin width is evaluated by repeating the fits with 2 (instead of 1)~\mev binning.
Additional uncertainties due to the background modelling are assigned by varying the spline function used to describe the WS distribution and by 
varying parameters of the AP polynomial function.

The relative efficiencies of the $B_2^*(5747)$ decays to $B^*\pi$ and $B\pi$ in each of the three \pt bins are fixed in the nominal fit.
These are varied independently to assign systematic uncertainties.
The uncertainties in the dependence of the efficiency on $Q$ value are propagated to the results by repeating the fit after varying, within their errors, the parameters of the polynomial function used to describe the variation.
Uncertainties are assigned for possible differences between data and simulation in the efficiency function by reweighting the simulation to match the \B momentum distributions observed in data.
Uncertainties are also assigned to take in account
the effect of changing the $\pt > 3 \gev$ cut on the \B candidate to $\pt > 4 \gev$, and of varying the boundaries of the three bins of the companion pion \pt.

Possible biases in the fits are investigated with ensembles of pseudoexperiments. No significant bias is found for most of the parameters, but shifts in the means and widths of the $B_J(5840)^0$ and $B_J(5960)^0$ states of up to 30\% of the statistical uncertainty are found and corrected for.
Systematic uncertainties corresponding to the size of the bias seen in the ensembles are assigned to all parameters.

Further systematic uncertainties are evaluated for the fixed fit parameters.
The spins of the higher mass states are changed from zero to two, the Blatt-Weisskopf effective radius is varied from its nominal value of $4 \gev^{-1}$ to $2$ and $6 \gev^{-1}$, and the $\Bstar-\B$ mass difference is varied within its uncertainty~\cite{PDG2014}.
The effects on the other parameters of the fit, when the $B_J(5840)^0$ and $B_J(5960)^0$ states are assumed to have natural spin-parity and hence contribute two peaks to the spectrum, are assigned as systematic uncertainties; the effects on the parameters of the $B_J(5840)^0$ and $B_J(5960)^0$ states themselves when changing this assumption are presented in Table~\ref{tab:naturalJPfitresults}.
Finally, the fits are repeated allowing for an additional state with a peak around $Q \sim 800 \mev$.
The additional state is not statistically significant, but the changes in the fitted parameters are assigned as systematic uncertainties.
The systematic uncertainties due to the momentum scale calibration are found
to be negligible.

In addition, various cross-checks are performed to ensure fit stability and reliability.
The stability of the data fits is studied by splitting the sample by the year of data taking, magnet polarity, and the charge of the companion pion.
The resulting independent samples are fitted using the same configuration as the nominal fit, and the results within each split are found to be consistent.

\section{Interpretation and conclusions}
\label{sec:interpretation}
The analysis of the invariant mass spectra of $\Bp\pim$ and $\Bz\pip$ combinations reconstructed with the \lhcb detector reported in this paper provides measurements of the properties of a number of \Bss resonant states.
The interpretation of the results is now given in two parts: firstly for the narrow \Bss signals, and subsequently for the broad, higher mass \Bss signals.

The narrow states are identified with the previously observed
$B_1(5721)^{0}$ and $B_2^*(5747)^{0}$ states, and their $B_1(5721)^{+}$ and $B_2^*(5747)^{+}$ isospin counterparts.
The peak positions in the $Q$-value distributions reported in Sec.~\ref{sec:nominalfit} can be converted into absolute masses using the known $B$ and $\pi$ meson masses and the $\Bstar-B$ mass difference~\cite{PDG2014}, leading to 
\begin{equation*}
\begin{array}{rccccccccccc}
\displaystyle m_{B_1(5721)^{0}}  & = & 5727.7\phani & \pm & 0.7\phani & \pm & 1.4\phani & \pm & 0.17 & \pm & 0.4 & \mev\,, \\
\displaystyle m_{B_2^*(5747)^{0}} & = & 5739.44 & \pm & 0.37 & \pm & 0.33 & \pm & 0.17 & & & \mev\,, \\
\displaystyle m_{B_1(5721)^{+}}  & = & 5725.1\phani & \pm & 1.8\phani & \pm & 3.1\phani & \pm & 0.17 & \pm & 0.4 & \mev\,, \\
\displaystyle m_{B_2^*(5747)^{+}} & = & 5737.20 & \pm & 0.72 & \pm & 0.40 & \pm & 0.17 & & & \mev\,, \\
\displaystyle \Gamma_{B_1(5721)^{0}} & = & \phani\phani30.1\phani & \pm & 1.5\phani & \pm & 3.5\phani & & & & & \mev\,, \\
\displaystyle \Gamma_{B_2^*(5747)^{0}} & = & \phani\phani24.5\phani & \pm & 1.0\phani & \pm & 1.5\phani & & & & & \mev\,, \\
\displaystyle \Gamma_{B_1(5721)^{+}} & = & \phani\phani29.1\phani & \pm & 3.6\phani & \pm & 4.3\phani & & & & & \mev\,, \\
\displaystyle \Gamma_{B_2^*(5747)^{+}} & = & \phani\phani23.6\phani & \pm & 2.0\phani & \pm & 2.1\phani & & & & & \mev\,.
\end{array}
\end{equation*}
The listed uncertainties are, from left to right: the statistical uncertainty, the experimental systematic uncertainty, and, where applicable,
the uncertainty on the \B meson mass and the uncertainty on the $\Bstar-B$ mass difference. Note that $B_1(5721)^{0,+}$ and $B_2^*(5747)^{0,+}$ notations are maintained here for consistency with the previous literature, even though the values of the masses no longer agree with these labels within uncertainty.
The results reported above are the most precise determinations of these quantities to date.

The relative branching fractions for the $B_2^*(5747)^{0,+}$ decays are measured to be
\begin{equation*}                                                                                                                                                                  
\begin{array}{rcccccc}                                                                                                                                                             
\displaystyle  \frac{{\cal B}\left( B_2^*(5747)^{0} \to \Bus\pim \right)}
  {{\cal B}\left( B_2^*(5747)^{0} \to \Bp\pim \right)} & = & 0.71 & \pm & 0.14
  & \pm & 0.30 \,, \\ \\ 
\displaystyle  \frac{{\cal B}\left( B_2^*(5747)^{+} \to \Bds\pip \right)}
  {{\cal B}\left( B_2^*(5747)^{+} \to \Bz\pip \right)} & = & 1.0\phani & \pm & 0.5\phani
  & \pm & 0.8\phani \,,                                                                                                                                                                  \end{array}                                                                                                                                                                        
\end{equation*}
where the uncertainties are statistical and systematic, respectively.
The significances of the  $B_2^*(5747)^{0,+} \to B^* \pi$ decays are evaluated using a
likelihood ratio test. 
Values of $6.5\sigma$ and $1.8\sigma$ are obtained for $\Bus\pim$ and
$\Bds\pip$, respectively, when only the statistical uncertainty is considered.
The inclusion of systematic uncertainties reduces the significance for the $\Bus\pim$ case to $3.7\sigma$. 
This result therefore corresponds to the first evidence for the $B_2^*(5747)^{0} \to \Bus\pim$ decay.
The relative branching fractions for the $B_2^*(5747)^{0,+}$ decays are in agreement with theoretical predictions~\cite{Orsland:1998de, Dai:1998ve, Zhong:2008kd, Godfrey:1986wj, Colangelo:2012xi}.

Structures at higher mass are clearly observed in the $Q$-value distributions.
To investigate the significance of the high mass states, large samples of
pseudoexperiments are generated and fitted with different configurations.
To cover the dominant systematic uncertainty on the yield of these states
which arises due to lack of knowledge of the shape of the AP component, the
pseudoexperiments are generated with the AP shape that minimises the
significance. 
A first ensemble is generated without any high mass states included.  Each
pseudoexperiment in this ensemble is fitted twice, once with the same model as used
for generation and once with an additional high mass resonance included.  
The distribution of the difference of \chisq values between the two fits is
extrapolated to obtain the $p$-value corresponding to the
probability to find a \chisq difference as large or larger than that obtained 
from the corresponding fits to data.
This procedure gives significances of $9.6\sigma$ for the $\Bp\pim$ case and
$4.8\sigma$ for the $\Bz\pip$ case.  

A second ensemble of pseudoexperiments is generated with a configuration that
corresponds to the best fit to the data with a single high mass resonance. 
The pseudoexperiments in this ensemble are fitted both with the model used for
generation and with a second high mass resonance included.  
The significances of the second peaks, again obtained from the difference in
\chisq values, are found to be $7.5\sigma$ and $4.6\sigma$ for the $\Bp\pim$
and $\Bz\pip$ cases, respectively. 
Since isospin symmetry is expected to hold for these states, 
this shows that under the hypothesis that the high mass structures are due to resonances, two new pairs of particles are observed.

Masses and widths of the $B_J(5840)^{0,+}$ and $B_J(5960)^{0,+}$ states are obtained with different
fit models, as discussed in Sec.~\ref{sec:Fitmodel}, and the corresponding
results are shown in Table~\ref{tab:B_5960_results}. 
The properties of the $B_J(5960)^{0,+}$ states are consistent with and more
precise than those obtained by the CDF collaboration when assuming decay to
$B\pi$~\cite{Aaltonen:2013yya}.
If the $B_J(5840)^{0,+}$ and $B_J(5960)^{0,+}$ states are considered under the quark model hypothesis, their properties are consistent with those expected for the $B(2{\rm S})$ and $B^*(2{\rm S})$ radially excited states.

\begin{table}[b]
  \centering
  \caption{
    Parameters of the $B_J(5840)^{0,+}$ and $B_J(5960)^{0,+}$ states obtained with different fit models. The empirical fit uses two, and the quark model fits three, RBW shapes to model the broad resonances.
    The listed uncertainties are, from left to right: the statistical uncertainty, the experimental systematic uncertainty, and, where applicable, the uncertainty on the \B meson mass and the uncertainty on the $\Bstar-B$ mass difference. 
    Note that any state not explicitly labelled as ``natural'' is considered to have unnatural spin-parity (and not to be $0^+$); the reported mass can be converted into the corresponding result under the $0^+$ spin-parity assumption by subtracting the $\Bstar-B$ mass difference.  
    Units of \mev are implied.}
  \label{tab:B_5960_results}
\begin{tabular}{r|ccccccccc}
\hline

                          & \multicolumn{9}{c}{\textrm{Empirical
                            model}} \\
\hline
$m_{B_J(5840)^{0}}$ &5862.9 & $\pm$ & \phani5.0 & $\pm$ & \phani6.7& $\pm$ & 0.2 & & \\
$\Gamma_{B_J(5840)^{0}}$ &\phani127.4 & $\pm$ & 16.7 & $\pm$ & 34.2 & & & & \\
$m_{B_J(5960)^{0}}$ &5969.2 & $\pm$ & \phani2.9 & $\pm$ & \phani5.1& $\pm$ & 0.2 & & \\
$\Gamma_{B_J(5960)^{0}}$ &\phani\phani82.3 & $\pm$ & \phani7.7 & $\pm$ & \phani9.4 & & & & \\
\hline
$m_{B_J(5840)^{+}}$ &5850.3 & $\pm$ & 12.7 & $\pm$ & 13.7& $\pm$ & 0.2 & & \\
$\Gamma_{B_J(5840)^{+}}$ &\phani224.4 & $\pm$ & 23.9 & $\pm$ & 79.8 & & & & \\
$m_{B_J(5960)^{+}}$ &5964.9 & $\pm$ & \phani4.1 & $\pm$ & \phani2.5& $\pm$ & 0.2 & & \\
$\Gamma_{B_J(5960)^{+}}$ &\phani\phani63.0 & $\pm$ & 14.5 & $\pm$ & 17.2 & & & & \\
\hline
\hline

                          & \multicolumn{9}{c}{\textrm{Quark
                            model, $B_J(5840)^{0,+}$ natural}} \\
\hline
$m_{B_J(5840)^{0}}$ &5889.7 & $\pm$ & 22.2 & $\pm$ & \phani6.7& $\pm$ & 0.2 & & \\
$\Gamma_{B_J(5840)^{0}}$ &\phani107.0 & $\pm$ & 19.6 & $\pm$ & 34.2 & & & & \\
$m_{B_J(5960)^{0}}$ &6015.9 & $\pm$ & \phani3.7 & $\pm$ & \phani5.1& $\pm$ & 0.2 & $\pm$ & 0.4 \\
$\Gamma_{B_J(5960)^{0}}$ &\phani\phani81.6 & $\pm$ & \phani9.9 & $\pm$ & \phani9.4 & & & & \\
\hline
$m_{B_J(5840)^{+}}$ &5874.5 & $\pm$ & 25.7 & $\pm$ & 13.7& $\pm$ & 0.2 & & \\
$\Gamma_{B_J(5840)^{+}}$ &\phani214.6 & $\pm$ & 26.7 & $\pm$ & 79.8 & & & & \\
$m_{B_J(5960)^{+}}$ &6010.6 & $\pm$ & \phani4.0 & $\pm$ & \phani2.5& $\pm$ & 0.2 & $\pm$ & 0.4 \\
$\Gamma_{B_J(5960)^{+}}$ &\phani\phani61.4 & $\pm$ & 14.5 & $\pm$ & 17.2 & & & & \\
\hline
\hline
                          & \multicolumn{9}{c}{\textrm{Quark
                            model, $B_J(5960)^{0,+}$ natural}} \\
\hline
$m_{B_J(5840)^{0}}$ &5907.8 & $\pm$ & \phani4.7 & $\pm$ & \phani6.7& $\pm$ & 0.2 & $\pm$ & 0.4 \\
$\Gamma_{B_J(5840)^{0}}$ &\phani119.4 & $\pm$ & 17.2 & $\pm$ & 34.2 & & & & \\
$m_{B_J(5960)^{0}}$ &5993.6 & $\pm$ & \phani6.4 & $\pm$ & \phani5.1& $\pm$ & 0.2 & & \\
$\Gamma_{B_J(5960)^{0}}$ &\phani\phani55.9 & $\pm$ & \phani6.6 & $\pm$ & \phani9.4 & & & & \\
\hline
$m_{B_J(5840)^{+}}$ &5889.3 & $\pm$ & 15.0 & $\pm$ & 13.7& $\pm$ & 0.2 & $\pm$ & 0.4 \\
$\Gamma_{B_J(5840)^{+}}$ &\phani229.3 & $\pm$ & 26.9 & $\pm$ & 79.8 & & & & \\
$m_{B_J(5960)^{+}}$ &5966.4 & $\pm$ & \phani4.5 & $\pm$ & \phani2.5& $\pm$ & 0.2 & & \\
$\Gamma_{B_J(5960)^{+}}$ &\phani\phani60.8 & $\pm$ & 14.0 & $\pm$ & 17.2 & & & & \\
\hline
\end{tabular}
\end{table}

In summary, the $\Bp\pim$ and $\Bz\pip$ invariant mass distributions obtained
from LHC $pp$ collision data recorded at centre-of-mass energies of 7 and 8~\tev, corresponding to an integrated luminosity of $3.0 \invfb$,
have been investigated in order to study excited \B mesons.
Precise measurements of the masses and widths of the $B_1(5721)^{0,+}$ and $B_2^*(5747)^{0,+}$ states are reported.
Evidence is found for the $B_2^*(5747)^{0} \to \Bus\pim$ decay.
Clear enhancements over background are observed in the mass range $5850$--$6000
\mev$ in both $\Bp\pim$ and $\Bz\pip$ combinations.
Fits to the data, accounting for the apparent enhanced production of the high
mass states in the high transverse momentum region, allow the parameters of
these states, labelled $B_J(5840)^{0,+}$ and $B_J(5960)^{0,+}$, to be determined
under different hypotheses for their quantum numbers. 

\section*{Acknowledgements}

 
\noindent We express our gratitude to our colleagues in the CERN
accelerator departments for the excellent performance of the LHC. We
thank the technical and administrative staff at the LHCb
institutes. We acknowledge support from CERN and from the national
agencies: CAPES, CNPq, FAPERJ and FINEP (Brazil); NSFC (China);
CNRS/IN2P3 (France); BMBF, DFG, HGF and MPG (Germany); INFN (Italy); 
FOM and NWO (The Netherlands); MNiSW and NCN (Poland); MEN/IFA (Romania); 
MinES and FANO (Russia); MinECo (Spain); SNSF and SER (Switzerland); 
NASU (Ukraine); STFC (United Kingdom); NSF (USA).
The Tier1 computing centres are supported by IN2P3 (France), KIT and BMBF 
(Germany), INFN (Italy), NWO and SURF (The Netherlands), PIC (Spain), GridPP 
(United Kingdom).
We are indebted to the communities behind the multiple open 
source software packages on which we depend. We are also thankful for the 
computing resources and the access to software R\&D tools provided by Yandex LLC (Russia).
Individual groups or members have received support from 
EPLANET, Marie Sk\l{}odowska-Curie Actions and ERC (European Union), 
Conseil g\'{e}n\'{e}ral de Haute-Savoie, Labex ENIGMASS and OCEVU, 
R\'{e}gion Auvergne (France), RFBR (Russia), XuntaGal and GENCAT (Spain), Royal Society and Royal
Commission for the Exhibition of 1851 (United Kingdom).

\appendix
\section{Covariance matrices} 
\label{app:cov_syst_err}

Tables~\ref{tab:cov:bp} and~\ref{tab:cov:bz} each show both statistical and systematic correlations between the main parameters of interest in the $\Bp\pim$ and $\Bz\pip$ fits, respectively.
In each Table, the masses and widths of the two broad states are seen to be heavily correlated with each other because they overlap, while the parameters of the narrow states are correlated because of the overlap between the $B_1(5721)^{0,+}$ state and the $B_2^*(5747)^{0,+}$ feed-down. 

\begin{table}[!b] 
  \centering
  \caption{
    Statistical (top) and systematic (bottom) covariance matrices of the nominal $\Bp\pim$ fit, where $\mu$ and $\Gamma$ stand for the mean and width respectively.
    The parameters related to the AP and WS shapes and the signal yields are suppressed for brevity. Units of \mev for $\mu$ and $\Gamma$ are implied.
  }
  \label{tab:cov:bp}

  \resizebox{0.99\textwidth}{!}{
  \begin{tabular}{lccccccccc}
          \hline
& \multicolumn{2}{c|}{$B_1(5721)^0$} & \multicolumn{3}{c|}{$B_2^*(5747)^0$} & \multicolumn{2}{c|}{$B_J(5840)^0$} & \multicolumn{2}{c}{$B_J(5960)^0$}\\
& $\mu$ & \multicolumn{1}{c|}{$\Gamma$} & BF ratio & $\mu$ & \multicolumn{1}{c|}{$\Gamma$} & $\mu$ &  \multicolumn{1}{c|}{$\Gamma$} &  $\mu$ & $\Gamma$ \\ 
\cline{2-10}
$B_1(5721)^0$ $\mu$      & $\phanm0.5$ &  &  &  &  &  &  &  &  \\
$B_1(5721)^0$ $\Gamma$   & $\phanm0.8$ & $\phanm2.3$ &  &  &  &  &  &  &  \\
$B_2^*(5747)^0$ BF ratio & $-0.1$      & $-0.1$      & $\phanm0.0$ &  &  &  &  &  &  \\
$B_2^*(5747)^0$ $\mu$    & $\phanm0.1$ & $\phanm0.1$ & $\phanm0.0$ & $\phanm0.1$ &  &  &  &  &  \\
$B_2^*(5747)^0$ $\Gamma$ & $-0.2$      & $-0.4$      & $\phanm0.0$ & $\phanm0.0$ & $\phanm1.0$ &  &  &  &  \\
$B_J(5840)^0$ $\mu$      & $\phanm0.0$ & $-0.4$      & $\phanm0.0$ & $\phanm0.0$ & $\phanm0.0$ & $\phanm24.5$ &  &  &  \\
$B_J(5840)^0$ $\Gamma$   & $-0.1$      & $\phanm2.0$ & $\phanm0.0$ & $-0.4$      & $-1.2$      & $\phanm23.1$      & 278.9 &  &  \\
$B_J(5960)^0$ $\mu$      & $\phanm0.0$ & $\phanm0.1$ & $\phanm0.0$ & $\phanm0.0$ & $\phanm0.0$ & $\phanm\phani7.4$ & $\phanm21.2$ & $\phanm\phani8.3$ &  \\
$B_J(5960)^0$ $\Gamma$   & $\phanm0.1$ & $\phanm0.6$ & $\phanm0.0$ & $\phanm0.1$ & $\phanm0.9$ & $-21.4$           & $-41.2$ & $-10.2$ & $\phani59.4$ \\
      \hline
      $B_1(5721)^0$ $\mu$& $\phanm1.9$ &             &                     &              &               &            &             &            &    \\
$B_1(5721)^0$ $\Gamma$   & $\phanm1.0$ & 12.2       &                     &              &               &            &             &            &    \\
$B_2^*(5747)^0$ BF ratio & $-0.1$      & $-0.1$       & $\phanm0.1$ &              &               &            &             &            &    \\
$B_2^*(5747)^0$ $\mu$    & $\phanm0.1$ & $\phanm0.1$  & $\phanm0.0$ & $\phanm0.1$ &               &            &             &            &    \\
$B_2^*(5747)^0$ $\Gamma$ & $-0.2$      & $-0.3$       & $\phanm0.0$ & $\phanm0.0$ & $\phanm2.2$          &            &             &            &    \\
$B_J(5840)^0$ $\mu$      & $\phanm0.1$ & $\phanm0.1$  & $\phanm0.0$ & $\phanm0.0$ & $\phanm0.0$   & $\phanm44.6$      &             &            &    \\
$B_J(5840)^0$ $\Gamma$   & $-0.2$      & $-0.4$       & $\phanm0.0$ & $\phanm0.0$ & $\phanm0.1$   & $\phanm\phani0.0$ & 1172        &            &    \\
$B_J(5960)^0$ $\mu$      & $\phanm0.0$ & $\phanm0.0$  & $\phanm0.0$ & $\phanm0.0$ & $\phanm0.0$   & $\phanm\phani0.0$ & $\phanm0.0$  &  $\phanm26.0$     &    \\
$B_J(5960)^0$ $\Gamma$   & $\phanm0.2$ & $\phanm0.3$  & $\phanm0.0$ & $\phanm0.0$ & $\phanm0.0$   & $\phanm\phani0.0$ & $-0.1$       &  $\phanm\phani0.0$        & $\phani88.6$\\
      \hline
  \end{tabular}
}
\end{table}

\begin{table}[!tb] 
  \centering
  \caption{
    Statistical (top) and systematic (bottom) covariance matrices of the nominal $\Bz\pip$ fit, where $\mu$ and $\Gamma$ stand for the mean and width respectively.
    The parameters related to the AP and WS shapes and the signal yields are suppressed for brevity. Units of \mev for $\mu$ and $\Gamma$ are implied.
  }
  \label{tab:cov:bz}

  \resizebox{0.99\textwidth}{!}{
  \begin{tabular}{lccccccccc}
          \hline
      & \multicolumn{2}{c|}{$B_1(5721)^+$} & \multicolumn{3}{c|}{$B_2^*(5747)^+$} & \multicolumn{2}{c|}{$B_J(5840)^+$} & \multicolumn{2}{c}{$B_J(5960)^+$}\\
      & $\mu$ & \multicolumn{1}{c|}{$\Gamma$} &  BF ratio & $\mu$ & \multicolumn{1}{c|}{$\Gamma$} &  $\mu$ &  \multicolumn{1}{c|}{$\Gamma$} & $\mu$ & $\Gamma$ \\ 
\cline{2-10}
$B_1(5721)^+$ $\mu$       & $\phanm3.3$ &  &  &  &  &  &  &  &  \\
$B_1(5721)^+$ $\Gamma$    & $\phanm5.0$ & 12.7 &  &  &  &  &  &  &  \\
$B_2^*(5747)^+$ BF ratio  & $-0.9$      & $-1.5$      & $\phanm0.3$ &  &  &  &  &  &  \\
$B_2^*(5747)^+$ $\mu$     & $\phanm0.4$ & $\phanm0.4$ & $-0.1$      & $\phanm0.5$ &  &  &  &  &  \\
$B_2^*(5747)^+$ $\Gamma$  & $-0.8$      & $-1.9$      & $\phanm0.2$ & $\phanm0.1$ & $\phanm4.0$ &  &  &  &  \\
$B_J(5840)^+$ $\mu$       & $\phanm0.5$ & $-3.2$      & $-0.1$      & $\phanm1.6$ & $\phanm8.8$ & 161.3 &  &  &  \\
$B_J(5840)^+$ $\Gamma$    & $\phanm2.2$ & $\phanm9.4$ & $-0.7$      & $-0.9$      & $-7.6$      & $-42.5$ & $\phanm571.2$ &  &  \\
$B_J(5960)^+$ $\mu$       & $\phanm0.1$ & $\phanm0.0$ & $\phanm0.0$ & $\phanm0.2$ & $\phanm1.0$ & $\phanm20.7$ & $\phani-7.8$ & $\phanm16.6$ &  \\
$B_J(5960)^+$ $\Gamma$    & $-0.3$      & $\phanm1.0$ & $\phanm0.0$ & $-0.4$      & $-2.0$      & $-95.8$ & $-107.4$ & $-22.4$ & 210.2 \\
    \hline
    $B_1(5721)^+$ $\mu$    & $\phanm9.6$ &             &                     &              &               &            &             &            &    \\
$B_1(5721)^+$ $\Gamma$     & $\phanm3.7$ & 18.3       &                     &              &               &            &             &            &    \\
$B_2^*(5747)^+$ BF ratio   & $-0.8$      & $-1.1$       & $\phanm0.6$          &              &               &            &             &            &    \\
$B_2^*(5747)^+$ $\mu$      & $\phanm0.2$ & $\phanm0.3$        & $-0.1$         & $\phanm0.2$         &               &            &             &            &    \\
$B_2^*(5747)^+$ $\Gamma$   & $-0.8$      & $-1.2$       & $\phanm0.2$          & $-0.1$        & $\phanm4.3$          &            &             &            &    \\
$B_J(5840)^+$ $\mu$        & $-0.2$      & $-0.3$       & $\phanm0.0$          & $\phanm0.0$   & $\phanm0.1$  & 187.7       &             &            &    \\
$B_J(5840)^+$ $\Gamma$     & $\phanm3.0$ & $\phanm4.3$        & $-0.9$         & $\phanm0.2$   & $-1.0$       & $-0.3$      & 6371        &            &    \\
$B_J(5960)^+$ $\mu$        & $\phanm0.0$ & $\phanm0.0$          & $\phanm0.0$  & $\phanm0.0$   & $\phanm0.0$  & $\phanm0.0$ & $\phanm0.0$ &  $\phanm\phani6.4$      &    \\
$B_J(5960)^+$ $\Gamma$     & $\phanm0.0$ & $\phanm0.0$       & $\phanm0.0$     & $\phanm0.0$   & $\phanm0.0$  & $-0.1$      & $\phanm0.0$ &  $\phanm\phani0.0$        & 295.2  \\
\hline
  \end{tabular}
}
\end{table}

\clearpage

\vspace{4ex} 

\setboolean{inbibliography}{true}
\bibliographystyle{LHCb}
\bibliography{main,LHCb-PAPER,LHCb-CONF,LHCb-DP}
\setboolean{inbibliography}{false}

\newpage

\clearpage

\centerline{\large\bf LHCb collaboration}
\begin{flushleft}
\small
R.~Aaij$^{41}$, 
B.~Adeva$^{37}$, 
M.~Adinolfi$^{46}$, 
A.~Affolder$^{52}$, 
Z.~Ajaltouni$^{5}$, 
S.~Akar$^{6}$, 
J.~Albrecht$^{9}$, 
F.~Alessio$^{38}$, 
M.~Alexander$^{51}$, 
S.~Ali$^{41}$
% ...

{\footnotesize \it
% ...
$ ^{5}$Clermont Universit\'{e}, Universit\'{e} Blaise Pascal, CNRS/IN2P3, LPC, Clermont-Ferrand, France\\
$ ^{6}$CPPM, Aix-Marseille Universit\'{e}, CNRS/IN2P3, Marseille, France\\
% ...
$ ^{9}$Fakult\"{a}t Physik, Technische Universit\"{a}t Dortmund, Dortmund, Germany\\
% ...
$ ^{37}$Universidad de Santiago de Compostela, Santiago de Compostela, Spain\\
$ ^{38}$European Organization for Nuclear Research (CERN), Geneva, Switzerland\\
% ...
$ ^{41}$Nikhef National Institute for Subatomic Physics, Amsterdam, The Netherlands\\
% ...
$ ^{46}$H.H. Wills Physics Laboratory, University of Bristol, Bristol, United Kingdom\\
% ...
$ ^{51}$School of Physics and Astronomy, University of Glasgow, Glasgow, United Kingdom\\
$ ^{52}$Oliver Lodge Laboratory, University of Liverpool, Liverpool, United Kingdom\\
% ...
}
\end{flushleft}

\end{document}